\documentclass[11pt]{article}%
\usepackage[iso-8859-7]{inputenc}
\usepackage{epsfig}
\usepackage{graphicx}
\usepackage{amsfonts}
\usepackage{amssymb}
\usepackage{amsthm}
\usepackage{amsmath}
\usepackage{multirow}
\usepackage{cite}
\usepackage{mathtools}%
\newcommand{\newc}{\newcommand}
\newc{\ra}{\rightarrow}
\newc{\lra}{\leftrightarrow}
\newc{\ov}{\overline}
\newc{\pa}{\partial}
\newc{\be}{\begin{equation}}
\newc{\ee}{\end{equation}}
\newc{\ba}{\begin{eqnarray}}
\newc{\ea}{\end{eqnarray}}
\newc{\n}{\nu}
\newc{\D}{\Delta}
\newc{\eps}{\epsilon}
\newc{\la}{\lambda}
\newc{\e}{\epsilon}
\newc{\nn}{\nonumber}

\csname @addtoreset\endcsname{equation}{section}

\usepackage{float}

\begin{document}

\hfill CERN-PH-TH/2013-174
 \vskip 2truecm
\vspace*{3cm}
\begin{center}
{ {\bf  Neutrino mass textures from F-theory}}\\
\vspace*{1cm} {\bf
I. Antoniadis$^{1,\,\flat }$,   G. K. Leontaris$^{2}$}\\
\vspace{4mm}
$^1$  Department of Physics, CERN Theory Division, \\
CH-1211, Geneva 23, Switzerland\vspace{1mm}\\
$^2$ Physics Department, Theory Division, Ioannina University, \\
GR-45110 Ioannina, Greece\vspace{1mm}\\
\end{center}

\vspace*{1cm}
\begin{center}
{\bf Abstract}
\end{center}
\noindent

Experimental data on the  neutrino mixing and masses strongly
suggest an underlying approximate symmetry of the relevant Yukawa
superpotential terms. Intensive phenomenological explorations during
the last decade indicate that permutation symmetries such as $S_4$,
$A_4$ and their subgroups, under certain assumptions and vacuum
alignments, predict neutrino mass textures compatible with such
data. Motivated by these findings, in the present work we  analyse
the neutrino properties  in F-theory GUT models derived in the
framework of the maximal underlying $E_8$ symmetry in the elliptic
fibration. More specifically, we consider  local F-$SU(5)$ GUT
models and study in detail   spectral cover  geometries with
monodromies associated to the finite symmetries  $S_4, A_4$ and
their transitive subgroups, including the dihedral group $D_4$ and
${Z}_2\times {Z}_2$. We discuss various issues that emerge in the
implementation of  $S_4, A_4$ neutrino models in the F-theory
context and suggest how these can be resolved. Realistic models are
presented for the case of monodromies based on their transitive
subgroups.  We exemplify  this procedure with  a  detailed analysis
performed for the case of ${ Z}_2\times {Z}_2$ model.

\vfill $^{\flat}$ {\it On leave from CPHT (UMR CNRS 7644) Ecole
Polytechnique, F-91128 Palaiseau, France.}

    \newpage

\section{Introduction}

F-theory Grand Unified Models
(F-GUTs)\cite{Beasley:2008dc,Donagi:2008kj,Beasley:2008kw,Donagi:2008ca,Donagi:2009ra}
 provide novel ways to compute Yukawa interactions which are
capable of describing convincingly   the observed fermion mass
hierarchy. F-GUTs are associated to  seven branes wrapping a complex
surface $S$ in an elliptically fibered eight dimensional internal
space~\footnote{For reviews on F-GUT model building
see~\cite{Heckman:2010bq,Weigand:2010wm,Leontaris:2012mh,Maharana:2012tu}.
}. The precise gauge group is determined by the specific structure
of the singular fibers over the compact surface $S$. Since $E_8$ is
the highest symmetry of the elliptic fibration, the  gauge symmetry
of the effective model can in principle be any of the $E_8$
subgroups. Choosing $SU(5)_{GUT}$, one can in principle avoid exotic
matter in the spectrum~\cite{Beasley:2008kw}. The gauge symmetry can
be broken by turning on appropriate  fluxes~\cite{Donagi:2008kj}
 which at the same time generate chirality for matter fields.

In these constructions matter fields are represented by
wavefunctions  on the intersections of seven branes with  $S$. These
intersections are  two dimensional compact  Riemann surfaces known
as matter curves, along which, gauge symmetry is enhanced.   When
three  such matter  curves intersect at a  point in the internal
manifold, the symmetry is further augmented while a trilineal Yukawa
coupling is formed.  In  viable F-theory models the top Yukawa
coupling usually arises from a trilinear superpotential term.  Its
strength is then equal
 to the properly normalised integral of the product of the three overlapping wavefunctions corresponding to
 the states residing on the relevant matter curves.
 Two mechanisms have been suggested to obtain fermion mass hierarchy.  If all
families reside on the same matter curve,  one way is to introduce
non-commutative fluxes and take into account non-perturbative
effects~\cite{Cecotti:2009zf,Font:2009gq,Font:2012wq}. If however,
fermion generations are distributed on different matter curves,
hierarchy might emerge from a generalisation of the Froggat-Nilsen
mechanism. It is likely of  course that both mechanisms operate in
particular constructions.

In  F-theory models the implementation of the generalised
Froggat-Nilsen mechanism  is realised through the various discrete
or abelian symmetries which appear in a natural way in any F-GUT
construction\cite{Dudas:2009hu,King:2010mq,Dudas:2010zb,Leontaris:2010zd,
Callaghan:2011jj,Camara:2011nj,Antoniadis:2012yk,Pawelczyk:2013tza}.
In the present work we  focus on  $SU(5)_{GUT}$ gauge group, so that
such symmetries arise from the decomposition of $SU(5)_{\perp}$ i.e,
the commutant  with respect to  initial $E_8$. GUT representations
appear as bifundamentals under the $E_8\to SU(5)_{GUT}\times
SU(5)_{\perp}$ breaking, and consequently the matter multiplets are
characterised by the $SU(5)_{\perp}$'s residual symmetry.  This
final  symmetry escorting the $SU(5)_{GUT}$ is the one left intact
by certain monodromy actions which usually appear in F-theory
compactifications.  Such symmetries are extremely useful when -under
certain circumstances-  are used to  implement the r\^ole of family
symmetries. If properly chosen,  they can provide the model with   a
hierarchical  fermion mass spectrum and at the same time eliminate
dangerous superpotential terms.  Given the details of the internal
manifold and the F-theory compactification, the specific nature of
the residual symmetry might be any continuous or finite subgroup of
$ SU(5)_{\perp}$. As such, these can be the transitive groups of
$S_4$, i.e. $A_4,D_4, Z_4$ etc, or abelian symmetries.

Within  the above context, the last few years a considerable amount
of work has been devoted to the study of the fermion mass hierarchy
problem. Intensive studies focused on  F-derived GUT models
 employing  monodromies leading to discrete symmetries such as  $Z_2$ and $Z_3$.  These
 symmetries result to identifications of several matter curves a fact that could be adequate
 to allow for the existence of a tree-level Yukawa coupling -to be associated with the top quark- and
 predict a promising hierarchical charged fermion mass spectrum. However,
  we know that accumulating  neutrino data reveal a rather different  picture of the neutrino
  sector; this is characterised  by large mixing
  angles~\footnote{The literature on this subject is vast.  For some relevant work
  and reviews see~\cite{deMedeirosVarzielas:2006fc}.}
   and apparent mild neutrino hierarchy as opposed
  to the large hierarchies and relatively small mixing of the quark sector.
More than a decade ago, it has been observed that experimental
neutrino data could be satisfactorily described  in terms of charged
and neutral lepton  mass matrices which exhibit a particularly
elegant structure.
 In this simple scheme, the mixing  matrix takes the form
\ba
 V_{TB}&=&V^{\dagger}_{l}V_{\nu}= \left(
\begin{array}{lll}
 -\sqrt{\frac{2}{3}} & \frac{1}{\sqrt{3}} & 0 \\
 \frac{1}{\sqrt{6}} & \frac{1}{\sqrt{3}} & -\frac{1}{\sqrt{2}} \\
 \frac{1}{\sqrt{6}} & \frac{1}{\sqrt{3}} & \frac{1}{\sqrt{2}}
\end{array}
\right)\label{TBmx}
 \ea
 This structure is in accordance with
$\sin^2\theta_{12}=1/3$, (for an approximate value $\theta_{12}\sim
\pi/5$), and $\sin^2\theta_{23}=1/2$, (i.e. for
$\theta_{23}=\pi/4$), while the third angle is exactly zero
$\theta_{13}=0$. The above
 is called Tri-Bi (TB) maximal mixing~\cite{deMedeirosVarzielas:2006fc}. In a basis where the
charged lepton mass matrix is diagonal the most general  neutrino
mass texture
 diagonalised by the  TB-mixing matrix is
 \be
 m_{\nu}=
\left(
\begin{array}{lll}
 u & v & v \\
 v & u+w & v-w \\
 v & v-w & u+w
\end{array}
\right) \label{mnTB}
\ee
  These observations suggested that there might be an  invariance of the
 relevant Yukawa terms under some underlying  symmetry involving finite groups such us $S_4,A_4$ etc.
The  ensuing  decade,  accumulating experimental data have shown
deviations from TB-mixing and that $\theta_{13}$ is not exactly
zero.  Nevertheless, such small departures from the exact symmetry
can be related to various sources of symmetry breaking contributions
such as radiative corrections etc.

Discrete symmetries have been found to play significant r\^ole in
model building.  They naturally emerge in string derived effective
models and have been studied in various
applications~\cite{Ibanez:1991hv}.  However, non-abelian discrete
symmetries have only been recently investigated in a string
context~\cite{BerasaluceGonzalez:2012vb}.
 As we have mentioned above,  they also arise naturally in the context of F-theory
 however, up to know little has been done
 to investigate their implications in model building\cite{Marsano:2009gv}.

Motivated by the apparent success of discrete  groups in neutrino
mixing, in this work we attempt to build models which incorporate
monodromic actions entailing such finite symmetries. In particular
we will rely on the $S_4$ and its transitive subgroups. In doing so,
we will employ the techniques developed in the context of the
spectral cover~\cite{Donagi:2009ra}. In particular, adopting the
geometrical interpretation which associates the GUT symmetry to a
divisor $S$ of the elliptically fibered internal manifold, we can
use spectral covers to describe physics  in its  vicinity and
extract useful properties of the model. In the elliptic fibration
this can be described by the Tate model~\cite{Tate75}  and in the
$SU(5)$ case, the spectral cover ${\cal C}_5$ is associated to a
five degree polynomial whose coefficients encode  useful information
for the effective model. A natural way to  attain symmetries such as
$S_4, A_4, D_4, { Z}_2\times { Z}_2$ etc, dictated by the neutrino
sector, is to split the spectral cover to ${\cal C}_5\to {\cal
C}_4\times {\cal C}_1$. We will see however, that  the combined
GUT-symmetry and internal geometry restrictions leave little room
for a successful implementation of the larger finite groups such as
$S_4$ and $A_4$. Notwithstanding these problems, such symmetries deserve a thorough
study since one might evade many difficulties when the F-GUT
symmetry is replaced by that of the F-theory Standard
Model~\cite{Choi:2010nf}. We leave such an analysis for future work
whilst in the second part of the present paper we  deal with a more
realistic model which is based on the ${ Z}_2\times { Z}_2$ family
symmetry.

The paper is organised as follows. In section 2 we present the
origin of the various discrete symmetries in F-theory while in
section 3 we analyse in detail those emerging in $SU(5)_{GUT}$
models in the context of the spectral cover. In section 4 we analyse
their implications on the neutrino sector in several  effective
models. In section 5 we present our conclusions.  Finally in the
Appendix we give  details of the calculations.

\section{The Origin of Family Symmetries in F-theory }

As we have argued in the introduction, an attractive feature of F-theory models is the
 appearance  of discrete symmetries which, among other consequences,  govern the
 structure of the Yukawa sector of the effective low energy theory.  We are particulary interested
 in non-abelian discrete symmetries which are quite appealing when dealing with the neutrino sector.
 The implementation of such a scenario in the effective model  requires the study of the spectral
 cover construction and the description of the relevant monodromies.
 The most important issue however  is the determination of the conditions on the manifold and fluxes
related to the part of the spectral cover associated to
the non-Abelian  discrete gauge group. As we have already pointed out, the study of the local
geometry is conveniently described by introducing the spectral surface    ${\cal C}_5$.  Guided by the
neutrino phenomenology we will split this to   ${\cal C}_5\to   {\cal C}_4\times {\cal C}_1$.  The above
splitting associates the  ${\cal C}_4$  part with a fourth degree polynomial and
the ${\cal C}_1$ with the linear piece which corresponds to a $U(1)$.   According to the well
known procedure form type IIB-theories,  if we turn on  a flux
along $U(1)$  we can induce chirality on the  spectrum  -transformed under $SU(5)_{GUT}$
and the remaining symmetry of the spectral cover-  which
is an essential step to build a viable model.  It may also eliminate the colour triplet parts of the
Higgs fiveplets which, had they remained in the spectrum, they would lead to proton decay
at unacceptable rates.  On the other hand
the ${\cal C}_4$ part corresponding to the quartic polynomial
induces a monodromy group which is a transitive subgroup of the
 non-abelian discrete symmetry  $S_4$.  To decide which
 the transitive subgroup is, this requires further  knowledge of  the structure of  non-abelian fluxes
as well as investigation of  the topological properties of the coefficients  of the associated quartic polynomial.
In this paper we do not deal with the first issue which is a rather involved task.  Nevertheless,
from the point of view of the  low energy field theory model and its phenomenological implications that we are
examining in this work, it suffices to analyse the properties of the polynomial coefficients which
constitute non-trivial sections and give all the necessary information to  determine the  matter  spectrum.
 A second issue concerns the action of geometric symmetries on the matter wavefunctions.
 This is  important because, even in the absence of the colour triplets in the presence of flux effects, there are still
 problematic Yukawa couplings that can be avoided only when matter parity is defined. We will
 explain how $Z_2$ symmetries of geometric origin can have such an effect. In doing so,
  we will follow~\cite{Hayashi:2009bt} and argue that the manifold and flux data may turn to invariance
  properties of  the  spectral surface  and  see how this can translate
 into an action on the matter wavefunctions.

\subsection{Monodromies and discrete symmetries}

In the case of the elliptic fibration, the highest symmetry obtained
is  ${ E}_8$, with  all  matter fields embedded  in  its adjoint
representation. The most familiar models of F-theory origin are
based either on the $SU(5)_{GUT}$ or  $SO(10)_{GUT}$ groups. The
first class in particular  arises under the following breaking
pattern
\[ E_8\to SU(5)_{GUT}\times SU(5)_{\perp}\]
All $SU(5)_{GUT}$ matter representations have also transformation
properties under the second group factor, denoted here as
$SU(5)_{\perp}$, comprising the  properties of the normal bundle.
The decomposition of the adjoint representation of $ E_8$ under
$SU(5)_{GUT}\times SU(5)_{\perp}$ is
\[248 = (24, 1) + (1, 24) + (10,5) + (\bar 5, 10) + (\overline{10}, \bar 5) + (5,\overline{10})\]
If we appeal to the geometric origin of these symmetries,   we
associate the fiber singularity to $SU(5)_{GUT}$  and interpret the
$ SU(5)_{\perp}$ as the  group describing the  bundle in the
vicinity.

The representations containing the low energy fields reside on
matter curves dubbed here $\Sigma_{10}, \Sigma_{5}$ and
characterised by the  $SU(5)_{\perp}$ weights $t_i$. In general on
the two distinct kinds of matter curves we may have
\be
\begin{split}
\Sigma_{10}&: n_{10}\times 10_{t_i}+ \bar n_{\bar{10}}\times \overline{10}_{-t_i}\\
\Sigma_{5}&:  n_{5}\times \bar 5_{t_i+t_j}+  \bar n_{\bar 5}\times
5_{-t_i-t_j}
\end{split}
\ee where the integers $n_i$ count the number of corresponding
representations  which live on a particular matter curve. The
required chirality of the fermion spectrum is generated when  $n_i
\ne \bar n_i$ and this happens when fluxes are turned on along
appropriate directions inside $SU(5)_{\perp}$.

One would expect that  $ SU(5)_{\perp}$ reduces to four $U(1)$'s
which would play the r\^ole of flavour symmetries in the formation
of the Yukawa couplings.  A more careful analysis shows that there
is a variety of possibilities. In general one expects the existence
of a non-trivial subgroup of the Weyl group $S_5$ of  $
SU(5)_{\perp}$.

 A short description of the situations goes as follows: The roots of
$SU(5)_{\perp}$  satisfy the spectral cover equation
\be
{\cal C}_5: \;  b_0s^5+b_2s^3+b_3s^2+b_4s+b_5=0\label{SC5}
 \ee
The coefficient $b_1$ does not appear in the equation (\ref{SC5}),
because it represents the sum of the $SU(5)$ roots which is
identically zero. We denote $t_i$ the roots of (\ref{SC5}) and write
\be {\cal C}_5:\; b_0 \prod_{i=1}^5 (s-t_i)=0\label{C5A} \ee Clearly
the $b_k$ coefficients  and $t_i$'s are related with the well known
relations $b_5/b_0=t_1t_2t_3t_4t_5,\cdots$  etc, when coefficients
of the same power of $s$ are compared. Such relations will be
exploited in the subsequence.

This equation describes the geometric nature of a local patch around
the $SU(5)_{GUT}$ singularity. Then, the coefficients $b_k$ (assumed
to belong in some field ${\cal F}$) carry the information of the
geometry. However, to obtain the roots  $t_i$  requires inversion of
the equations  $b_k(t_i)$ and some solutions may not belong to the
field ${\cal F}$ but to some  ${\cal K}\supset {\cal F}$ ( the so
called splitting field ${\cal K}$). This means that the
corresponding symmetry may not be $U(1)$. Depending on the number of
roots which lie outside ${\cal F}$,  and the specific properties of
the coefficients, the group maybe any subgroup of $S_5$, namely
\[ S_4, A_4, Z_n\times Z_m, n+m<5,\;\dots\]
Such symmetries  are very  important since they determine several
properties of
 the matter curves and the spectrum of the model hosted on them.

\subsection{Spectral cover discrete symmetries}

A convenient way to prevent unwanted and dangerous terms in the Yukawa
Lagrangian is the implementation of a discrete ($Z_N$) symmetry.  Indeed, in most
effective low energy supersymmetric models matter parity is usually associated to a $Z_2$
symmetry, so that dangerous proton decay Yukawa interactions are avoided.  It has been
shown~\cite{Hayashi:2009bt} that such a  symmetry in F-theory can have a geometric origin associated
to the compactified space and the fluxes. Once such a transformation is identified it must be
checked whether its action  induces  the appropriate matter
parity on the   wavefunctions of the matter fields residing on the relevant  matter curves.
  Thus, for example one can assume a $Z_2$  background
 configuration  and try to relate this symmetry with the local geometry. To start from
 a global geometry however, it is not an easy task but one can assume its existence
 in certain compactifications and work out the details locally near the
 GUT divisor.  Under these assumptions a realisation is presented for a GUT divisor $S_{GUT}=\mathbb{P}^2$
 and a $Z_2$ transformation map in reference~\cite{Hayashi:2009bt}.  Locally, the manifold and the flux data
 can be captured by the Higgs bundle on $S_{GUT}$ which in turn can be described using the notion
 of the spectral surface and the line bundles.  In this context, it can be shown that
  a $Z_2$ symmetry in particular induces an $SU(3)$ rotation of the
three-complex coordinates of the total space which act  on spinors the same way.
A simple bottom-up approach how to incarnate such discrete  symmetries in
local models was applied in\cite{Antoniadis:2012yk} and could be described as follows.
 In F-theory, we  may
appeal to the geometric origin of the GUT symmetries and consider
the latter as a divisor $S_{GUT}$   which locally can be  covered by
open patches $U_a\in S_{GUT}$ ~\cite{Hayashi:2009bt}. For simplicity, here we
 focus on a single trivialisation patch  and take $s$ to be the coordinate along
the fiber and demand  that under the required geometric
transformation the spectral cover equation  should remain invariant
up to an overall phase. To this end consider the transformation
$\sigma$ where $s, b_k$ are mapped according to~\footnote{$Z_2$ transformations (on Enriques
$S_{GUT}$ broken with Wilson lines)
in conjunction with the Higgs bundle spectral cover have also been used subsequently
in~\cite{Marsano:2012yc} in global $SU(5)$ constructions.}
\be
s(\sigma(p))\;=\; s(p)\,e^{i\phi},\; b_k(\sigma(p))\;=\;
b_k(p)\,e^{i\xi} e^{-i(6-k)\phi}\label{phases} \ee Then each term in
the spectral cover equation transforms the same way \be
 b_k s^{5-k}\ra e^{i\xi}b_k s^{5-k}\label{specphase}
 \ee

We can exploit this invariance to communicate a $Z_N$ symmetry to
$S_{GUT}$. We take the phase $\phi$ to be
\[\phi=\frac{2\pi}{N}\]
where $N$ is an integer.  Thus, choosing $N=2$ for example, we have
$\phi=\pi$ and a $Z_2$ symmetry   is realised with $s,b_k$
transforming according to
\ba s\ra -s,\; b_k\ra (-1)^{k}
e^{i\xi}\,b_k\label{bphase}
\ea
while in the limit $N\to \infty$, we
get $s\to s$ and $b_k\to e^{i\xi}b_k, \forall k$. Such symmetries
can be communicated~\cite{Antoniadis:2012yk} to the
 matter curves as well as to the representations accommodated on them, imposing
 restrictions on the possible interaction and Yukawa terms of the superpotential.

\section{Non abelian discrete symmetries and fermion mass textures}

The issue of the fermion mass hierarchical structure has been
tackled in several recent works in the context of F-theory.  The
large hierarchy among the charged fermions (quarks and leptons) and
the CKM mixing
 can be  readily  accommodated either by non commutative flux effects or
 by employing the Frogatt-Nielsen mechanism interpreting the
  surplus $U(1)$'s as family symmetries. Solutions for the large neutrino
  mixing  have  also been considered. In the $SU(5)$ case, right-handed neutrinos
  are identified as Kaluza-Klein modes~\cite{Antoniadis:2002qm,Bouchard:2009bu}.
  Yukawa couplings including such
  states generate   a milder hierarchy and allow for a large mixing
  as required by the neutrino data.
The fermion mass hierarchy issues have been considered mainly in
cases of the $Z_2$ monodromies. In the present work, we pay
particular attention to the neutrino sector. Because much
phenomenological work  has been devoted to interpret the large
lepton mixing with the implementation of discrete non-abelian
groups, we intend to explore such cases in the context of F-theory
where these symmetries arise naturally. Successful finite groups
reconciling the neutrino data include the permutation symmetries
such as $S_4$ as well as some of its subgroups. Motivated by these
approximate symmetries of the neutrino sector, in the next section
we will  consider the case of the ${\cal C}_4\times {\cal C}_1$
 splitting of the spectral cover and attempt to derive the
peculiar neutrino properties, i.e. the large mixing and the tiny
neutrino mass scale.

\subsection{  ${\cal C}_4\times {\cal C}_1$ spectral cover in  $SU(5)$ }

We will assume $SU(5)$ unified models with $SU(5)_{\perp}$ spectral
covers.  Finite symmetries such as  $S_4, A_4$ and their subgroups
can be obtained directly by splitting the spectral cover according
to  ${\cal C}_5\to {\cal C}_4\times {\cal C}_1$, which implies  the
splitting of the polynomial as \be \sum b_k s^{5-k} =
(a_1+a_2s+a_3s^2+a_4s^3+a_5s^4) (a_6+a_7s)
 \label{C4C1}
 \ee
defined in terms of new coefficients $a_i$. The  splitting
(\ref{C4C1})  induces the breaking of   $SU(5)_{\perp}$ to
 a Galois group (which is $ S_4$ or some other  subgroup) and a $U(1)$.
 More precisely,
 depending on the specific properties of  $a_{1...5}$ coefficients the
 discrete group could be one of $S_4,  A_4,  D_4,  V_4={ Z}_2\times { Z}_2,  Z_2$.
 We can deduce  the topological properties  of these
 coefficients by exploiting the relations $b_k(a_i)$. These can be found by
comparing coefficients of the same powers in (\ref{C4C1}). One finds
the following relations
 \be
  b_0=a_5a_7,\;  b_5=a_1a_6,\; {\rm and}\;
  b_k=a_{6-k}a_6+a_{5-k}a_7,\;{\rm for}\; k=1,2,3,4\,
\label{bsas0}
 \ee
 We can readily find that the homologies satisfy relations of the form
 \[ [b_k]=[a_l]+[a_{12-l-k}]\]
 for all  combinations of indices appearing in (\ref{bsas0}).
 The $b_k$ topological  properties are well defined and can be expressed in terms of
 the first Chern classes $c_1$ and $-t$  of the tangent and the normal bundle  respectively.
Thus, given that\cite{Donagi:2008ca}
\be
[b_k]= \eta -k \,c_1,\; \eta= 6c_1-t\label{homol}
\ee
all $[a_i]$ can be expressed in terms of one arbitrary parameter
taken here to be $[a_6]=\chi$. These are presented in
Table~\ref{T0}.
\begin{table}
\begin{center}
\begin{tabular}{|c|c|c|c|}
\hline $a_k (k=1,\dots,5)$ &$a_6$&$a_7$&$a_0$
\\
\hline $\eta-(6-k)c_1-\chi$& $\chi$& $c_1+\chi$&$\eta-2c_1-2\chi$
\\
\hline
\end{tabular}
\end{center}
\caption{Homology classes for coefficients $a_i$  for the ${\cal
C}_4\times {\cal C}_1$ ($SU(5)$) case} \label{T0}
\end{table}

To proceed further with the restrictions  we impose  the
tracelessness condition $b_1=\sum_{i=1}^5 t_i=0$ of $SU(5)_{\perp}$.
The corresponding equation in (\ref{bsas0}) reads
 \[
a_5a_6+a_4a_7=0\] and is solved
 by introducing a suitable coefficient  $a_0$
\be a_4= \pm\, a_0 a_6,\;\;\; \; a_5= \mp\, a_0 a_7\label{b1cond}
\ee

We discuss now the implications of these topological properties on
the matter curves. The $SU(5)_{GUT}$ fiveplets  transform as the
$10\in SU(5)_{\perp}$ thus, we have
 a bifundamental  $(5,\,10_{\perp}) $. Denoting with $t_i$ the $SU(5)_{\perp}$ weights,
the $5/\bar 5's \in SU(5)_{GUT}$   are discriminated by the $\frac
12 5(5-1)=10$ sheets of the spectral cover, with the assignments
\[ t_i+t_j, \; \; i<j=1,\dots, 5\]
 These are `roots' of  the relevant equation\cite{Donagi:2009ra}
  \[ {\cal P}_{5}(s)\propto \prod_{i<j}(s-t_i-t_j)= s^{10}+\cdots +c_1\,s+c_0\]
  The $c_0=\prod_{i,j}(t_i+t_j)$ is a symmetric quantity thus, it can be expressed in terms
  of the $b_k$ coefficients. Consequently, the defining  equation $P_5={\cal P}_5(0)$ for
  the GUT fiveplets is computed to be~\cite{Donagi:2009ra}
\ba
 P_5&=&b_4 b_3^2-b_2 b_5 b_3+b_0 b_5^2\label{P5s}
 \ea
  Hence, in principle there
could be ten distinct curves (Riemann surfaces) accommodating the
fiveplets, but the actual number equals  the number of factors $P_5$
splits to. Plugging in the $a_k$ conditions enforced by the equation
$b_1=0$ we see that this splits to two  factors
\[ P_5= P_5^a P_5^b=
\left( a_2^2a_7+ a_2a_3 a_6\mp a_0 a_1 a_6^2\right) \left(a_3 a_6^2+
\left(a_2 a_6+a_1  a_7\right)a_7\right)
\]
For the upper and lower signs in $b_1=0$ solution respectively.

 If no further assumptions are made,
this means that the 10 curves accommodating the fiveplets reduce to
two distinct ones, with homologies
 readily computed from Table~\ref{T0}
\[ [P^a_5]= 2[a_2]+[a_7]=2\eta-7 c_1-\chi,\;\;\;\; [P^b_5]=[a_3]+2[a_6]=\eta-3c_1+\chi\]

The $SU(5)_{GUT}$ tenplets  transform  in $5\in SU(5)_{\perp}$,
therefore
 are associated to the equation of the spectral cover
 of the fundamental representation. In the case of splitting the spectral cover
 as in (\ref{C4C1}), we get
\be P_{10}={\cal C}_5(0)=a_1a_6\label{10s}
\ee
This equation defines two  matter curves characterised by the
homologies  of  $a_1,a_6$.

The invariance of the spectral cover equation under the phase
transformations (\ref{specphase}) can also equip the matter curves
with an  additional transformation property which can be conveyed to
the corresponding representations. We may for example assume a $Z_2$
invariance of the spectral cover equation. Then, the most general
transformation  of the coefficients $a_i$ compatible with the
$b_k$'s properties under $Z_2$, are
\[ a_{1,3,5}\to e^{-i\varphi}a_{1,3,5},\;\; a_{2,4}\to -e^{-i\varphi},\;
a_6\to -e^{i\varphi}a_6,\; a_7\to e^{i\varphi}a_7\] An important
observation here is that although the parity of the spectral cover
is fixed to $Z_2$, there is still freedom to define a different
symmetry on the matter curves themselves. This freedom  is very
useful  particularly for the $A_4$ and $S_4$-neutrino models where
symmetries such as $Z_3,Z_4$ are often used to prevent unwanted
Yukawa terms~\cite{deMedeirosVarzielas:2006fc}.
   \begin{table}[t] \centering%
\begin{tabular}{|l|c|l|c|c|}
\hline
$SU(5)$& Equation &Homology&$M_i$&$Z_{n}$\\
\hline $
10^{(1)}_{t_i}$&$a_1$&$\eta-5c_1-\chi$&$M_{10}$&$e^{-i\varphi_n}$\\
${10}^{(2)}_{t_5}$&$a_6$&$\chi$&$M_{10}'$&$-e^{i\varphi_n}$\\
$  5^{(0)}_{-2t_i}$&$a_2^2a_7+\dots$&$2\eta-7
c_1-\chi$&$M_5$&$e^{-i\varphi_n}$\\
$  \bar 5^{(1)}_{t_{i}+t_5}$&
$a_3a_6^2+\cdots$&$\eta-3c_1+\chi$&$M_5'$&$e^{i\varphi_n}$\\
 \hline
\end{tabular}%
\caption{Field representation content under $SU(5)$, the defining
equations and matter content  and parity is defined by
$\varphi_n=2\pi/n$  for the ${\cal C}_4\times {\cal C}_1$ case (the
indices $i,j,k$ take the values $1,2,3,4$).
 The multiplicities should satisfy $\sum M_i=0$.}
\label{SU5S4a}
\end{table}

We collect the various  properties of the representation content
residing on the various matter curves in Table~\ref{SU5S4a}.
 Bulk fields and singlets (not shown in this Table) eventually
  may appear   in the spectrum, and will be discussed later in the context of
  specific constructions.
In the last column a phase transformation has been assigned for each
representation, as a consequence of  the $Z_2$ invariance of the
spectral cover equation. Choosing the phase $\varphi=\pi$ we obtain
the standard matter parity.

\subsection{Discrete Monodromies}

\begin{figure}[!b]
\centering
\includegraphics[scale=0.53]{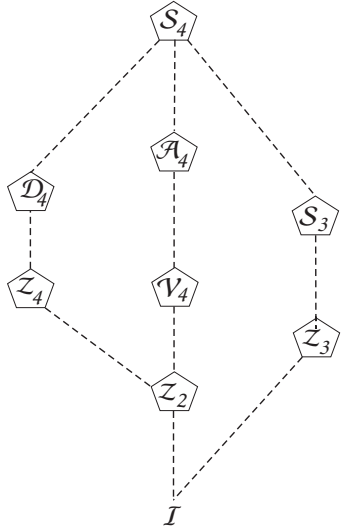}
\caption{Basic $S_4$ subgroup structure. $S_n$ is the permutation
symmetry of $n$ objects. $D_4$ is the dihedral group, while $V_4\sim
Z_2\times Z_2$ is the Klein group.} \label{S4sub}
\end{figure}

The particular choice of monodromy  plays a decisive r\^ole for the
viability of the model since it regulates many of the parameters of
the effective theory, including mass matrices and mixing. Our
present choice of the  ${\cal C}_4\times {\cal C}_1$ splitting has
been dictated by the peculiar features of the neutrino sector which,
in many phenomenological explorations, have been correlated to $S_4$
and its subgroups. In this section, we  pursue further the
investigation to these cases in the F-theory context.  In the
present construction, this chain of discrete symmetries is
associated to the ${\cal C}_4$ spectral cover part with
corresponding equation: \be {\cal C}_4: \; {\cal P}_4(s)=
a_1+a_2s+a_3s^2+a_4s^3+a_5s^4=0\label{C4cover} \ee This is an
irreducible polynomial whenever the roots lie outside the field
${\cal F}$ where the coefficients belong to. The minimal extension
${\cal K}$ of the field ${\cal F}$  which  contains the roots,
defines the splitting field of ${\cal P}_4$. Depending on the
specific properties of $a_i$ the remaining symmetry group might be
any of the $S_4$ subgroups and in particular
\[ S_4, A_4, D_4, Z_4, S_3, V_4= Z_2\times Z_2, Z_3, Z_2\]
For later convenience, we depict the basic subgroup chains of $S_4$
in figure~\ref{S4sub}.

In order to   determine which of the above is the Galois group, we
will  examine partially symmetric polynomials  of the roots of the
corresponding equation (\ref{C4cover}). If we denote with $t_i$ the
roots of the above polynomial, we know that any symmetric polynomial
$f(t_1,\dots,t_4) $  can be written in terms of the quantities
$s_1=\sum t_i=-a_4/a_5, \dots, s_4=t_1t_2t_3t_4=a_1/a_5$ in a unique
way. A partially symmetric function is left invariant only under a
specific subgroup of $S_4$ and this defines the Galois group.

\subsection{The $A_4$ case}

In this section we will introduce appropriate partially  symmetric functions of roots $t_i$
which remain invariant only under the desired $S_4$  subgroups.  Then, for each particular
case we will investigate the implications for the  coefficients $a_i$.

 We start the analysis with the discriminant  $\Delta(t_i)$
 of the quartic polynomial which is of particular importance.
 This is a symmetric function of the roots $t_i$, written as follows
 \[ \Delta=\prod_{i\ne j} (t_i-t_j)=\prod_{i< j} (t_i-t_j)^2\]
 and as such, remains invariant under the whole $S_4$ group.
Now, we recall first that the $A_4$ discrete subgroup
 involves only the even permutations of $S_4$.
 If we take the quantity
\[\delta = (t_1-t_2) (t_1-t_3)(t_1-t_4)(t_2-t_3)(t_2-t_4)(t_3-t_4)\]
we can readily observe that this is invariant only under even
permutations of roots. Observe now that $\delta$ is the square root
of the discriminant of the polynomial related to ${\cal C}_4$
\[\Delta(t_i)=
\delta^2(t_i)\]

We proceed now by examining the latter  as functions of the
coefficients $a_k$. As a symmetric function, the discriminant
 $\Delta$  is given  in terms of the polynomial
coefficients and is proportional to the Sylvester determinant given
by
\be S_d=\left|
\begin{array}{lllllll}
 a_5 & a_4 & a_3 & a_2 & a_1 & 0 & 0 \\
 0 & a_5 & a_4 & a_3 & a_2 & a_1 & 0 \\
 0 & 0 & a_5 & a_4 & a_3 & a_2 & a_1 \\
 4 a_5 & 3 a_4 & 2 a_3 & a_2 & 0 & 0 & 0 \\
 0 & 4 a_5 & 3 a_4 & 2 a_3 & a_2 & 0 & 0 \\
 0 & 0 & 4 a_5 & 3 a_4 & 2 a_3 & a_2 & 0 \\
 0 & 0 & 0 & 4 a_5 & 3 a_4 & 2 a_3 & a_2
\end{array}
\right| \label{Disc0}
\ee
 while $S_d= a_5\Delta$.
 Substituting the $b_1=0$ conditions found in (\ref{b1cond})
 for the upper signs, $S_d$ takes the form
   \be
   \label{Disc1}
\begin{split}
S_d=& a_7  a_0^2 \left(27 a_7^2 a_0 a_2^4+2 a_6 a_0 \left(2 a_0
   a_6^2+9 a_3 a_7\right) a_2^3\right.\\
  & \left.-\left(4 a_7 a_3^3+\left(a_3 a_6^2+144 a_1
   a_7^2\right) a_0 a_3+6 a_1 a_6^2 a_7 a_0^2\right) a_2^2\right.
   \\
   &\left.
   +2 a_1 a_6
   a_0 \left(-40 a_7 a_3^2-9 a_6^2 a_0 a_3+96 a_1 a_7^2 a_0\right)
   a_2\right.\\
   &\left.+a_1 \left(16 a_7 a_3^4+4 \left(a_3 a_6^2+32 a_1 a_7^2\right) a_0
   a_3^2+27 a_1 a_6^4 a_0^3\right.\right.\\
   &+\left.\left.16 a_1 a_7 \left(9 a_3 a_6^2+16 a_1 a_7^2\right)
   a_0^2\right)\right)
   \end{split}
   \ee
  The form of the discriminant for the lower signs of (\ref{b1cond})
  is derived by just substituting $a_0\to -a_0$.
  For arbitrary  $a_i$'s  (more precisely imposing no other condition on $a_i$'s),
  in general    the discriminant cannot be written as a square of a quantity with all elements in
   ${\cal F}$.  Then, the Galois group is  $S_4$ or one of the subgroups $D_4,Z_4$.
   If the discriminant is a square of a quantity $\delta$ then
   the Galois group is    $A_4$ or $V_4=Z_2\times Z_2$.

      \subsection{ The  $\Delta =\delta^2$ constraints on $a_k$'s}

We will attempt to unravel  possible relations of the  $a_i$ coefficients  for the case of the $A_4$ symmetry.
   Undoubtedly, the relevant  condition $\Delta =\delta^2$  could be met only for   very particular
   relations among  $a_k$.
The search for any  correlations however, is not a trivial task,
given the complicated form of (\ref{Disc1}).

 We choose to proceed
with this investigation as follows: If we consider $\Delta$ as a polynomial
of a certain  $a_k$ coefficient ($k=1,...,5 $), i.e.,
\[  \Delta\to f(a_k)= c_n a_k^n+\cdots +c_0,\;\; n\le 4\]
a necessary condition for  $\Delta$ to be written as a square is the
vanishing of the  discriminant of the  polynomial $f(a_k)$. There
are 5 such coefficients, hence there are an equal number of choices.
To choose a suitable  case, we expand the discriminant in each of
the  available coefficients. We obtain
\be
\Delta=\left\{
\begin{array}{ll}
f_5&=0 \,{\bf a_5^4}+256 a_1^3 {\bf a_5^3}+\cdots
\\
f_4&=-27 a_1^2 {\bf a_4^4}+\cdots
\\
f_3&=16 a_1 a_5{\bf a_3^4 }+\cdots
\\
f_2&= -27  a_5^2{\bf a_2^4}+\cdots
\\
f_1&=0 \,{\bf a_1^4}+256 a_5^3{\bf  a_1^3}+\cdots
\end{array}\right.
\ee
From these  five expansions we see that in two cases ($f_1,f_5$),
$\Delta$ can be expressed as a third order polynomial whilst in the
three remaining cases it is a polynomial of  fourth degree. A necessary condition
to write the third degree polynomials  as a square is that the
coefficient of the highest (third) degree is positive definite, i.e.
${\rm sign}(a_1a_5)=(+)$. As far as the fourth degree ones are
concerned, this criterion is fulfilled   only for $f_3$, provided
the same condition holds,  i.e. ${\rm sign}(a_1a_5)=(+)$. The
highest degree coefficients of the other two
fourth-degree polynomials $f_2, f_4$ cannot be positive definite and as such they
are automatically rejected.

There is an advantage of the fourth degree polynomial since when the
required criteria are satisfied this can be a product of two second
degree ones, and it is possibly more convenient to handle them.

Under these  circumstances, the only case to satisfy the requiring
criteria is when we consider $\Delta $ as a polynomial with respect
to $a_3$ , i.e,
\[\Delta\to f_3 (a_3)=c_4 a_3^4+c_3 a_3^3+c_2a_3^2+c_1a_3+c_0\]
where all $c_k$  are known functions of  $a_i$.
Computing the discriminant of $f_3$ , we find that it is proportional to  $\Delta_{f_3}\propto D_a^2D_b^3$ with
\be
\begin{split}
D_a&=a_1 a_4^2-a_2^2 a_5\\
D_b&=\left(27 a_1^2 a_4-a_2^3\right) a_4^3-6 a_1 a_2^2 a_5 a_4^2+4096 a_1^3 a_5^3+3
   a_2 \left(9 a_2^3-256 a_1^2 a_4\right) a_5^2
   \end{split}
   \ee
   As we have pointed out, a necessary condition to write $\Delta =\delta^2$ is  the vanishing of the
    discriminant $\Delta_{f_3}$  of the polynomial $f_3$.
 There are two ways to satisfy this constraint, either by setting $D_a=0$ or for $D_b=0$.  With a detailed discussion given
 in the Appendix, we record here the  induced conditions among the coefficients. From $D_a=0$ we readily see that
 \be
a_1 a_4^2=a_2^2 a_5\label{Dacon}
\ee
As explained in the Appendix, this condition is not sufficient to write $\Delta =\delta^2$. This happens if
we require
\be
a_2^2=2\,a_1\,a_3,   \; a_4^2=2\,a_3\,a_5\label{DAzero}
\ee

From the second $D_b=0$, after several algebraic manipulations we end up with the condition
\ba
( a_2^2a_5-a_4^2a_1)^2&=&\left(\frac{16a_1a_5-a_2a_4}{3}\right)^3\label{Dbcon0}
\ea
Plugging in the above the $SU(5)$ constraint $b_1=0$ yields
\ba
( a_2^2a_7-a_0a_1a_6^2)^2&=&a_0\left(\frac{16a_1a_7+a_2a_6}{3}\right)^3\label{Dbcon}
\ea

As it is expected, such relations will have also implications on the matter curves. Thus, from the first condition in particular we find that
the fiveplets split to three orbits according to
\be
 P_5=a_6(2a_0a_1a_6-a_2a_3)  \left(a_3 a_6^2+
\left(a_2 a_6+a_1  a_7\right)a_7\right)\label{5A4}
\ee
with their homologies computed directly from those of $a_i$'s given
in Table ~\ref{T0}.

    \subsubsection{The   spectrum   of the $SU(5)\times A_4\times U(1)$ model}

In this section we present the $SU(5)$ spectrum subject to the
constraints of the $ A_4$ monodromy.  In order to clarify  its
properties we find it convenient to change the notation with respect
to $t_{1,\dots,4}$-roots in  an $A_4$ basis. To this end, we recall
that the permutations of the four roots $t_{1,2,3,4}$ constitute the
$S_4$ group while their sum
\[t_s= t_1+t_2+t_3+t_4\]
clearly remains invariant under all permutations. This is orthogonal
to the three remaining combinations
\[ t_a=t_1+t_2-t_3-t_4,\; t_b=t_1-t_2+t_3-t_4,\; t_c=t_1-t_2-t_3+t_4\]
which form an $S_4$-triplet. The action of only  $S_4$-even
permutations of roots  ( $Q_{even}$) , corresponds to the $A_4$
subgroup. Under these permutations we   see that
\[ Q_{even} t_{\gamma} =\pm t_{\delta} ,\; (\gamma,\delta )\in (a,b,c)\]
From   equations (\ref{10s})  of the tenplets we can see that those
corresponding to the four roots $t_{1,2,3,4}$ are associated  to
$a_1$. These can be expressed as a singlet and a triplet under
$S_4/A_4$ as follows
\[ \Sigma_{10_1}:\;(10,1) =10_{t_s},\;    \Sigma_{10_3}:\;(10,3)=\{10_{t_a},10_{t_b},10_{t_c}\}\]
This means that $P_{10}$ should split to three factors which
essentially  requires  $a_1$ to be the product of two factors
$a_1\sim \beta_0\beta_1$.

We turn now our attention to the fiveplets. In the new basis
$t_{s,a,b,c}$, the $\bar 5_{t_i+t_j}$'s are written as  $t_s\pm
t_{\gamma}$, with $\gamma=a,b,c$
and  similarly for the $ 5$'s defined by    $-(t_i+t_j)$ . Thus,
this subset of  $SU(5)$ multiplets  form  two  triplets under $A_4$
, namely
\[ 3_{\pm} =\frac 12 (t_s\pm t_{a,b,c})\]
   whose components are interchanged by the action of $A_4$,   i.e., $Q_{even} : 3_+ \to 3_-$.
   The corresponding matter curves  accommodate the  elements
   \[ (5,3)_{(\pm)} =(5_{\frac 12 (t_s\pm t_{a})},\,5_{\frac 12 (t_s\pm t_{b})},\,5_{\frac 12 (t_s\pm t_{c})})\]

In the new `basis'    the  elements $5_{t_i+t_5}$ form a singlet and
a triplet, i.e.
\[ 5_{t_1+t_5}\to 5_{\frac 14(t_s+t_a+t_b+t_c)+t_5},\]
and
\[ ( 5_{t_2+t_5},  5_{t_3+t_5},  5_{t_4+t_5})\to ( 5_{\frac 14(t_s+t_a-t_b-t_c)+t_5}, 5_{\frac 14(t_s-t_a+t_b-t_c)+t_5},5_{\frac 14(t_s-t_a-t_b+t_c)+t_5})\]


We also need to rewrite the  $t_i-t_j$ differences in the new basis,
 for the classification of the $SU(5)$ singlets. These can be expressed as  $t_i-t_j=t_{\gamma}-t_{\delta}$
 for ${\gamma,\delta}= a,b,c$.
Accordingly, for $t_i-t_5$ we have a singlet $\theta_0
=(1,1)_{-t_5}$ and a triplet $\theta_3=(1,3)_{-t_5}$ in analogy with
the fiveplets $\bar 5_{t_i+t_5}$ above. All representations are
collected and assigned under the new notation in Table \ref{SU5S4}.
\begin{table}[t] \centering%
\begin{tabular}{|rl|l|c|}
\hline
Curve&$SU(5)\times A_4\times U(1)$& $t_{\gamma}$&$matter$\\
\hline $\Sigma_{10_a}:$&$ F=(10,3)$&${t_a}$& $F_i=(Q,u^c,e^c)_i$
\\
 $\Sigma_{10_b}:$&$  F_x= (10,1)$&${t_s}$&\\
   $ \Sigma_{10_c} :$&$ F_y=(\ov{10},1)_{-t_5}$&${-t_5}$&\\
   $\Sigma_{5_1}:$&$  H=\bar (5,3)$&${t_s\pm t_a}$&$h_u$\\
     $\Sigma_{5_2}:$&$  \bar f= (\bar 5,3)_{+t_5}$&${\frac 14(t_s-t_a)+t_5}$&$ \bar
f_i=(\ell,d^c)_i$\\
$\Sigma_{5_3}:$&$  \bar H= ( \bar 5,1)_{+t_5}$&$\frac
14(t_s+3t_a)+t_5$&$\bar h_d $
\\
$\Sigma_{1_a}:$&$ \theta_a= (1,3)$&0&$\theta_{ij} $\\
  $\Sigma_{1_b}:$&$  \theta_b= ( 1,3)$&$t_a$&$\theta_{i4} $\\
   $\Sigma_{1_c}:$&$  \theta_c= ( 1,3)_{-t_5}$&${\frac 14(t_s-t_a)-t_5}$&$\theta_{i5} $
   \\
   $\Sigma_{1'}:$&$ \theta'= (1,1)_{-t_5}$&$-t_5$&$\theta_{45} $\\
    $   \Sigma_{1''}:$&$ \theta''= (1,1)_{+t_5}$&$t_5$&$\theta_{54} $
\\
$\Sigma_{1_0}:$&$  \theta= (1,1)$&$0$&$\theta$\\
\hline
\end{tabular}%
\caption{Field representation content for the $SU(5)\times A_4\times
U(1)$ case 1. For alternative models, matter resides on
$\Sigma_{10_{b,c}}$ curves  (see text).}\label{SU5S4}
\end{table}

\subsection{ Further reduction of the discrete group}

We will examine the conditions on $a_k$ coefficients to reduce the
$S_4/A_4$ down to their subgroups  (we follow closely the notations
of~\cite{Artin}). Given the property of the discriminant, (i.e
whether it is a square root or not) we may have a different symmetry
breaking chain. We have seen what we should expect an $A_4$ discrete
group (or its subgroup $V_4$) if $\Delta =\delta^2$. On the
contrary, if $\Delta \ne \delta^2$ we will see that
 the Galois group is $S_4$ or $D_4, Z_4$.
 To check this we need to examine other partially symmetric functions of roots.
To this end, consider  the sums~\footnote{See for example
~\cite{Artin} and reference~\cite{Marsano:2009gv}.}
\[x_1= t_1t_2+t_3t_4,\;  x_2=t_1t_3+t_2t_4,\;x_3=t_3t_2+t_1t_4,\]
To construct an appropriate function of $t_i$, we assume  the
associated cubic polynomial (the resolvent of ${\cal P}_4(s)$)  with
roots the $x_{1,2,3}$
\be
 f(x)=(x-x_1) (x-x_2) (x-x_3)\label{qubic}
 \ee
This is invariant under $S_4$ and has the same discriminant as
${\cal P}_4(s)$.

We examine here a partially symmetric quantity (function of roots)
which is invariant under the Dihedral subgroup $D_4$. This is the
sum
\[x_1= t_1t_2+t_3t_4\]
This remains unaltered by the following eight $S_4$ elements
\[ I, (12), (34), (12)(34), (14)(23), (13)(24), (1423), (1324)\]
This is one of the three dihedral subgroups $D_4$ of $S_4$.
Similarly we work for the other two quantities $x_{2,3}$ each
remaining invariant under the appropriate  dihedral subgroup $D_4$
(see table~\ref{S4D4}).
   \begin{table}[t] \centering%
\begin{tabular}{|l|c|l|}
\hline
$D_a$&$I, (12), (34), (12)(34), (14)(23), (13)(24), (1423), (1324)$&$t_1t_2+t_3t_4$\\ 
$D_b$&$I, (13), (24), (12)(34), (14)(23), (13)(24), (1234), (1432)$&$t_1t_3+t_2t_4$\\ 
$D_c$&$I, (14), (23), (12)(34), (14)(23), (13)(24), (1243),
(1342)$&$t_1t_4+t_2t_3$\\ \hline
\end{tabular}%
\caption{The three Dihedral groups isomorphic to $D_4$ and their
invariants.} \label{S4D4}
\end{table}
The symmetry splits the fiveplets in three orbits according to
\be
 R_a =\{t_1+t_2,\,t_3+t_4\},\;  R_b =\{t_1+t_3,\,t_2+t_4\},\;  R_c =\{t_1+t_4,\,t_2+t_3\},
 \ee

 The  coefficients of $f(x)$ can be computed as functions of $a_k$. In
particular,  the $S_4$ invariant quantity
\[
f(0)= x_1x_2x_3=(t_1t_2+t_3t_4)( t_1t_3+t_2t_4)(t_3t_2+t_1t_4)
\]
is expressed in terms of $a_k$'s as follows
\be
 f(0)= 4a_5a_3a_1-a_4^2a_1-a_5a_2^2\label{fzero}
\ee
Substitution of the conditions for $b_1=0$ transform the above to
\[f(0)=\alpha _0 \left(a_2^2 a_7-a_1 \left(\alpha _0 a_6^2+4 a_3
   a_7\right)\right)\]
If the  polynomial is irreducible, then  $f(0)\ne 0$. This, and the
condition $\Delta\ne\delta^2 $ imply that the group is $S_4$.

Now let's examine the case of a reducible $f(x)$. The polynomial
$f(x)$ can  be factorised for $f(0)=0$
\[ f(x) =  x^3+f_1 x^2+f_2 x= x (x^2+f_1 x+f_2 )\]
This yields
\[a_2^2 a_7=a_1 \left(\alpha _0 a_6^2+4 a_3
   a_7\right)\]
   Substitution of the latter into the equation of fiveplets gives
   \[P_5 =a_3 \left(a_2 a_6+4 a_1 a_7\right) \left(a_3 a_6^2+a_7
   \left(a_2 a_6+a_1 a_7\right)\right)\]
Therefore, for the case of dihedral symmetry $D_4$  the fivelpets
split to three orbits with homologies
\[ [a_3] =\eta -3c_1-\chi,\;\;
[a_2 a_6]= \eta -4 c_1,\;\;  [a_2 a_6 a_7] =\eta- 3c_1+\chi\]

Finally, notice that the substitution of  equations (\ref{DAzero})
to (\ref{fzero}) automatically implies $f(0)=0$ thus these are just
the requirements for the symmetry reduction down to $V_4\sim
Z_2\times Z_2$.


The present analysis with respect to $\Delta$ and $f(x)$ is
summarised in Table~\ref{Galois}.

\begin{table}[t]
\centering%
\begin{tabular}{l|cc}
Discriminant& $f(x)$& Group\\
\hline
$\Delta\ne \delta^2$&  $f(0)\ne 0\;$: & $S_4$\\
  &  $f(0)= 0\;$: & $D_4$\\
  \hline
  $\Delta=\delta^2$&  $f(0)\ne 0\;$: & $A_4$\\
  &  $f(0)= 0\;$: & $V_4$\\
\end{tabular}%
\caption{The Galois groups for the various cases of the Discriminant
and the reducibility of the resolvent cubic
$f(x)=(x-x_1)(x-x_2)(x-x_3)$  in (\ref{qubic}).} \label{Galois}
\end{table}

\subsection{Embedding of $A_4$ models in the $SU(4)$ spectral cover}

Among other cases discussed in the previous sections,
 we have also derived the constraints on the coefficients $a_i$ for which
 the monodromy group reduces to $A_4$.  In this case, we find it useful to determine the
embedding of the fields $10_{t_a}, \bar 5_{t_a+t_b}$ etc in $A_4$
representations.

We can proceed as follows:  having in mind  the particular splitting
of the spectral cover we are dealing with in this work, i.e. ${\cal
C}_4\times {\cal C}_1$, for our present purposes we may consider the
$SU(4)$ as the covering group of the monodromy and the embedding of
the fields into  its $4$ and $6$  representations. Indeed, recall
that  ${ E}_8\supset SO(10)\times SU(4)$  while
 matter resides in  the ${ E}_8$ adjoint representation 248,
which in this case decomposes as follows
\begin{center}
$ { 248}\;\ra\;(45,1)+(\overline{16}, { \bar 4_{\perp}})+
$\fbox{$(16, {4_{\perp}})+(10, {6_{\perp}})+(1,  { 15_{\perp}})$}
\end{center}
The relevant -w.r.t.  ordinary matter- representations are included
in the box. $16's$ and $10's$ include the non-trivial $SU(5)$
representations while $(1,  { 15_{\perp}})$ is an $SU(5)_{GUT}$
singlet.

As demonstrated in~\cite{Luhn:2007sy},  the $A_4$ representations
are in one-to-one correspondence with those of $SU(3)$, thus the
$SU(4)$ ones decompose  according to the pattern shown in
Table~\ref{43A4}.
   \begin{table}[t] \centering%
\begin{tabular}{|l|c|l|}
\hline
$SU(4)$& $SU(3)$&$A_4$\\
\hline
$4$&$3+1$&$3+1$\\ 
$6$&$3+\bar 3$&$3$\\ 
$15$&$8+1+3+\bar 3$&$3+3'+1+1'+1''$\\ \hline
\end{tabular}%
\caption{Decomposition of the ${\cal C}_4$ spectral cover symmetry
to $A_4$} \label{43A4}
\end{table}
In this case the  spectrum emerges according to the chain rules \be
\begin{split}
(16,4)&\stackrel{\rm SO(10)\times SU(3)}
     {\longrightarrow} (16,3)+(16,1)\\
       &\stackrel{\rm SU(5)\times A_4}
     {\longrightarrow} [(10,3)+(\bar 5,3)+(1,3)]+[[(10,1)+(\bar 5,1)+(1,1)]\\
(10,6)&\stackrel{\rm SO(10)\times SU(3)}
     {\longrightarrow} (10,3)+(10,\bar 3)\\
      &\stackrel{\rm SU(5)\times A_4}
     {\longrightarrow} (5,3)+(\bar 5,3)\\
(1,15)&\stackrel{\rm SO(10)\times SU(3)}
     {\longrightarrow} (8+1+3+\bar 3)\\
     &\stackrel{\rm SU(5)\times A_4}
     {\longrightarrow} (1,3)+(1,3')+1+1'+1''
\end{split}
\ee These results are in accordance with the analysis of our
previous sections.  In the next section, we will work out a few
examples of effective models  paying  attention to the neutrino
sector.

\section{ Effective low energy theory}

In this  section we  seek to build effective low energy models which
incorporate the discrete symmetries analysed above. Motivated by the
successful implementation of the $A_4$ in the neutrino sector, we
will mainly focus in this case  and its subgroups.

\subsection{Neutrino masses from $A_4$}

Observing the spectrum presented in Table~\ref{SU5S4}, we notice
that there are more than one ways to accommodate the generations on
matter curves. We present these possibilities starting with the
charged sector.

\subsubsection{Case 1}

Taking the point of view that three of the tenplets are accommodated
as an $A_4$-triplet, as a first example, we choose to accommodate
the three families of quark doublets etc, in $(10,3)$ since in this
context it is the only possible way to obtain a tree-level top-quark
coupling. We notice however that in this case it is not  obvious how
to generate chirality for the $SU(5)$ representations.  In this case
we envisage that in the yet unspecified {\it global} geometrical
structure, higher dimensional non-Abelian internal fluxes will
restrict non-trivially on the $\Sigma_{(10,3)}$ matter curve
inducing a chiral spectrum sitting in $(10,3)$. There is a second
shortcoming in this picture which will be revealed as soon as we
write down the charged fermion mass terms.

We  start by making  the assignment for the remaining SM fermions
and Higgs according to Table~\ref{SU5S4}. Then the following
couplings emerge
\be
\begin{split}
{\cal W}_u &\supset \; (10,3)_{t_i}\,(10,3)_{t_i}\, (5,3)_{-2t_i}\\
{\cal W}_{d,\ell} &\supset \; (10,3)_{t_i}\,(\bar 5,3)_{t_i+t_5}\, (\bar 5,1)_{t_4+t_5}\theta_{i5}\\
{\cal W}_{\nu} &\supset \; (\bar 5,3)_{t_i+t_5}\,(\bar
5,3)_{t_i+t_5}\, ( 5,3)_{-2t_i}
\, ( 5,3)_{-2t_i}\theta_{i5}\theta_{i5}\\
\end{split}
\ee
Recall that the $A_4$ tensor products are
\[3\times 3\to 1+1'+1''+3_s+3_a,\; 1'\times 1''\to 1\]

Notice that the up-Higgs fiveplet is also an $A_4$ triplet. The most
general vev can be written
\[ h_u =(h_1,h_2,h_3),\;\to\;\langle h_u\rangle =(v_1,v_2,v_3)\]
For the up quarks we get
\be
\begin{split}
F\,F\,h_u & = (2F_1F_1-F_2F_3-F_3F_2)\,v_1\\
&  +(2F_3F_3-F_1F_2-F_2F_1)\,v_3\\
&  +(2F_2F_2-F_1F_3-F_3F_1)\,v_2\\
\end{split}
\ee
Then, to first order, the  matrix  for the up quarks is
\[
m_{u}=\left(
\begin{array}
[c]{ccc}%
2v_1 & -v_3 & -v_2\\
-v_3& 2v_2&-v_1\\
-v_2&-v_1&2v_3
\end{array}
\right)
\]
For the  leptons one gets
\be
\begin{split}
F\,\bar f\, \bar H\,\theta_{i5} &=(10,3)_{t_i}(\bar
5,3)_{t_i+t_5}(\bar 5,1)_{t_4+t_5}\theta_{i5}
                  \end{split}
\ee The $SU(5)$ singlets $\theta_{ab}$ participating in this
coupling  compose an $A_4$-triplet. Assuming the triplet vev
$\langle\theta_{ab}\rangle\sim (a_1,a_2,a_3)$, in this case we get
\[ m_{\ell}\sim \left(
\begin{array}{lll}
 2 a_1  &-a_3  & -a_2 \\
 -a_3  & 2 a_2 & -a_1 \\
 -a_2 & -a_1& 2a_3
\end{array}
\right)\, \langle h_d\rangle\] The down quarks and charged leptons
look identical, however, in general  radiate effects  are expected
to discriminate them.

We observe that there is a generic structure of tree-level masses
for the charged fermions' sector. Hence, in both cases, the
eigenvalues are given by  common formulae, satisfying the sum rule
\[m_3= m_1+{m_2}=\sum_i a_i\]
Apparently, this cannot be satisfied by any mass relations of quarks
and/or charged leptons. Higher order corrections cannot make this
compatible with data, therefore, the present assignment  appears to
be  too restrictive. We proceed to an alternative scenario.

\subsubsection{ Case 2}
To avoid the problem with  the charged fermion sector  we assume
that the 10 representation of $SU(5)$  accommodating the left handed
doublet quarks of the three families is an $A_4$ singlet.
 We  may use either of the two matter curves $\Sigma_{10_{b,c}}$ to
 accommodate the three families. For example, if we assign
\[  F_x =(10,1) = (Q,d^c e^c)\]
then, for the up quarks the allowed coupling is  the following
fourth order one
\[\lambda_t\frac{1}{\Lambda}\, (10,1) \,(10,1)\,(5,3) \,(1,3)\leftrightarrow
\lambda_t\frac{1}{\Lambda} \,F_x F_x H \theta_b \] Assigning
$\langle H\rangle=(v_{1},v_2,v_3), \langle
\theta_b\rangle=(b_{1},b_2,b_3)$, this induces a mass for the third
generation up quark
\[   m_t\sim \lambda_t \frac{v_1b_1+v_2b_3+b_3v_2}{\Lambda} \]
If  all families are on the same matter curve,  according
to~\cite{Cecotti:2009zf,Font:2009gq}  the lighter generations
receive masses from non-commutative fluxes and instanton effects.

For down quarks and leptons we can write a fifth order coupling
\[\frac{1}{\Lambda^2}(10,1) \,(\bar 5,3)_{t_5}\,(\bar 5,1)_{t_5} \,(1,3)_{-t_5}^2\leftrightarrow
 \frac{1}{\Lambda^2}F_x f \bar H \theta_c^2\]
with a  mass eigenvalue associated to the bottom quark
\[m_b\sim \lambda_b\frac{\langle \theta_c^2\rangle}{\Lambda^2}\,\langle h_d\rangle\]
In this scenario, non-perturbative  corrections and non-commutative
fluxes are expected to generate the lighter masses.

As an alternative possibility, we mention that families may also
reside on $\Sigma_{10_c}$. The corresponding representations
$10_{t_5}$ are charged under the $U(1)_{t_5}$ factor. As a
consequence,  chirality is readily obtained by turning on
appropriate $U(1)$ flux along $t_5$, however the disadvantage here
is that the top Yukawa coupling arises only at sixth order.

\subsubsection{Neutrinos}

We turn now our attention to  the couplings of the neutrinos. There
is an effective Majorana operator at sixth order
\[{\cal W}_{\nu}= \bar f\cdot\bar  f\,\cdot \bar H\cdot\bar H\,\cdot \theta_{i5}\cdot\theta_{i5}\]
Inasmuch the involved fields are triplets under the $A_4$ symmetry,
 there are various ways to obtain $A_4$ invariants.
Notice first
\be
\begin{split}
\bar f\cdot\bar  f\to 3\times 3& =1_{f}+1_{f}'+1_{f}''+3_{f}+3_{f}'\\
\bar H\cdot\bar H \to 3\times 3& =1_h+1_h'+1_h''+3_h+3_h'\\
\theta_{i5}\cdot\theta_{i5}\to 3\times 3&=
1_{\theta}+1_{\theta}'+1_{\theta}''+3_{\theta}+3_{\theta}'
\end{split}
\ee Hence, we have several $A_4$ invariants  contributing to
$m_{\nu}$ such as
\be
\begin{split}
m_{\nu}&\propto 1_{\theta} ( 1_f1_h +1_f'1_h''+1_f''1_h')\\
&+1_{\theta}'  ( 1_f1_h'' +1_f''1_h+1_f'1_h')\\
&+1_{\theta}''  ( 1_f1_h' +1_f''1_h+1_f''1_h'')\\
&+1_{\theta} \cdot (3_f\times 3_h)_1+\cdots
\end{split}
\ee

With the above analysis at hand, we proceed now to a simple example.
We would like to see whether at a first approximation the specified
neutrino mass structure is in accordance with large mixing as
indicated by the experimental data. Therefore, we take a toy example
and choose the vev alignments to be
\[\left\{a_1\to 1,a_2\to 0,a_3\to 0,v_1\to 0,v_3\to v_2\right\}\]
If we assign with $V_{\ell}$ the unitary matrix diagonalising the
charged lepton sector and $V_{\nu}$ the corresponding one for the
neutrino mass matrix, the lepton mixing matrix to be compared with
the experiment is the product $V=V_{\ell}^{\dagger}V_{\nu}$. In the
case of the charged leptonic sector discussed previously, the
structure of the mass matrix is generated by non-perturbative
contributions and  the mixing effects are expected to be small. In
this case the large mixing effects measured in the neutrino
experiments are generated by  $V_{\nu}$. In the present model, the
neutrino mass  matrix takes the form
\[m_{\nu}\propto \left(
\begin{array}{lll}
 2 & 1 \cdot c & 1 \cdot  c \\
 1\cdot c& 13 & - 4\cdot c \\
 1\cdot c & -  4\cdot c & 13
\end{array}
\right)
\]
In the above texture we have inserted an arbitrary coefficient $c$
to parametrise corrections from the charged lepton sector and
renormalisation effects. This is a simplified approximation, however
at present we would like to show that the observed large mixing
effects arise indeed from the neutral sector of the theory. Hence,
the choice $c=2$ yields \be V_{\nu}=\left(
\begin{array}{lll}
 -0.85689 & 0.515499 & 0. \\
 0.364513 & 0.605913 & -0.707107 \\
 0.364513 & 0.605913 & 0.707107
\end{array}
\right)
\ee
 Although in this  simplified procedure the prediction for the $\theta_{13}$
  mixing angle turns out to be strictly zero, even small charged
  lepton mixing will lift this to a non-zero value, hopefully
  compatible with data. Undoubtedly,  it is remarkable that
  in general the mixing angles are pretty close to the measured values.
The eigenmasses are
\[ V_{\nu}^{\dagger}m_{\nu}V_{\nu}\;\to \;\left(
 0.3 ,  6.7 , 21.0
\right)\,m_0
\]
predicting a ratio $r=\Delta m_{atm}^2/\Delta m_{sun}^2\approx 10$.
This value is not far from from the experimental one ($r\sim 30$).
Nevertheless, a more detailed investigation should take into account
renormalisation group effects which are beyond the purpose of the
present work.

\subsection{ ${ Z}_2\times {Z}_2$ Models }

 In recent attempts to interpret the neutrino data in the
 context of field theory models, a wide number of
 other (simpler) discrete symmetries have also been proposed.
 Motivated by these attempts
  in this section we will  examine neutrino mass textures which  are
  derived  only from combinations of ${ Z}_n$ symmetries.
 Based on the analysis of previous works on F-theory models, here
 we will  derive  the neutrino mass textures
  for the  ${ Z}_2\times { Z}_2$ case. In conjunction with
  our previous analysis, this may arise under the following braking
  chain
  \[ SU(5)_{\perp}\to SU(4)\to A_4\to {Z}_2\times { Z}_2\]

   \begin{table}[t] \centering%
\begin{tabular}{|l|c|c|c|c|}
\hline
$ SU(5)$&$U(1)$ ``weight"& homology & $U(1)_Y$&$U(1)_X$\\
\hline $10^{(1)}$& $\pm {t_{1,2}}$&
$\eta-2c_1-{\chi}-\psi$&$ N_1$ &$M_{10}^1$\\
$ 10^{(2)}$&$\pm {t_3}$&  $-2c_1+\chi$&$ N_2$
&$M_{10}^2$\\
$ 5^{(0)}$&$\mp ({t_{1}+t_2})$&  $-c_1+{\chi}+\psi$&$ -N_1$ &$M_{5}^0$\\
 $ 5^{(1)}$&$\mp ({t_{1,2}+t_3})$&  $2\eta -4c_1-2{\chi}-2\psi$&$ 2N_1$
&$M_5^1$\\
$  5^{(2)}$&$\mp ({t_{1,2}+t_5})$&$\eta
-2c_1-{\chi}-\psi$&$ N_1$ &$M_5^2$\\
$5^{(3)}$&$ \mp({t_{3,4}+t_5})$& $ -2c_1+{\chi}+2\psi$&$
-2N_1-N_2$ &$M_5^3$\\
  $  5^{(4)}$&$\mp ({t_{3}+t_4})$& $-c_1+\chi$&$N_2$&$M_5^4$\\
 $ 10^{{(3)}}$&$\pm {t_{5}}$& $-c_1+\psi$&$-N_1-N_2$&$M_{10}^3$\\ \hline
\end{tabular}%
\caption{ $SU(5)$ matter curves, their homology class and the
$U(1)_{X,Y}$ flux parameters   for the ${ Z}_2\times { Z}_2$ case.
The signs $\pm t_i$ are associated to $10/\ov{10}$ representation
accommodated on the specific matter curve, while $\mp ( t_i+t_j)$ to
$5/\bar 5$ ones.} \label{SU5Z2Z2}
\end{table}

\subsubsection{$SU(5)$ spectrum}

In the $SU(5)$ model with ${ Z}_2\times { Z}_2$ discrete symmetry,
the spectral cover equation for $SU(5)_{\perp}$ is written in the
form \ba {\cal C}_{5}(s)&=&  \left(a_3 s^2+a_2 s+a_1\right)
\left(a_6 s^2+a_5 s+a_4\right) (a_7+a_8 s)\label{c5z2z2} \ea
Proceeding as above, we start by identifying the relations
$b_k(a_i), k=1,\dots 5$ by comparing coefficients of
 the same power in $s$ of the polynomials
(\ref{SC5}) and (\ref{c5z2z2}).

The $10\in SU(5)$ are in one to one correspondence with the
solutions of the equation
\[b_5= a_1a_4a_7=0\]
 therefore they are associated to $a_1=0, a_4=0$
and $a_7=0$. The fiveplets are found  by solving  the corresponding
equation (\ref{P5s}) which  in terms of the $a_i$'s in the present
case  can be written~\cite{Dudas:2010zb,Antoniadis:2012yk}:
 \be
\label{Z2Z2}
\begin{split}
P_5&= (a_6 a_7 + a_5 a_8) \times (a_1^2 - a_1 (a_5 a_7 + 2 a_4 a_8)
\lambda +
   a_4 (a_6 a_7^2 + a_8 (a_5 a_7 + a_4 a_8)) \lambda^2)\\
&\times (a_1 -
   a_5 a_7 \lambda) \times (a_6 a_7^2 + a_8 (a_5 a_7 + a_4 a_8))
  \times a_5  \end{split}
   \ee
   This equation has  five factors corresponding to an equal number of fiveplets
  dubbed as \[5^{(0)},5^{(1)},5^{(2)},5^{(3)},5^{(4)}\]
    in the order of appearance  in the  product (\ref{Z2Z2}).  It is straightforward to determine
    their  homology classes,  while to compute
    the flux     restrictions we define
  \[{\cal F}_Y\cdot {\psi}=-N_1-N_2,\;
  {\cal F}_Y\cdot {\chi}=N_2,\; {\cal F}_Y\cdot \eta ={\cal F}_Y\cdot c_1=0\]
  where $N_{1,2}$ integers and ${\cal F}_Y$ the hypercharge flux.

  The results are summarised in Table \ref{SU5Z2Z2}. In the first
  column we write the $SU(5)$ representation with a superscript
  denoting the specific matter curve. In the second column we write the
  corresponding $SU(5)_{\perp}$ root. The third column shows the
  homology classes of the matter curves computed
  in~\cite{Dudas:2010zb,Antoniadis:2012yk}. The last two columns
  show the integers $N_i, M_j$  associated to hypercharge and
  $U(1)_X$ fluxes respectively. The latter are subject to the
  restriction  $\sum_j M_j=0$~\cite{Dudas:2010zb}.

We proceed with a specific example, based on the particular choice
of fluxes  $N_1=0, N_2=1$.  For a given matter curve $\Sigma_{10}$
with $M_{10}, N_Y$ units of $U(1)_X\in SU(5)_{\perp}$ and $U(1)_Y$
fluxes correspondingly, the  multiplicities of the resulting { SM}
spectrum are given by
 \ba {\bf 10}\in
SU(5)\Rightarrow\left\{\begin{array}{lll}
n_{(3,2)_{\frac 16}}-n_{(\bar 3,2)_{-\frac 16}}&=&{ M_{10}}\\
n_{(\bar 3,1)_{-\frac 23}}-n_{(3,1)_{\frac 23}}&=&{ M_{10}}-{ N_Y}\\
n_{(1,1)_{1}}-n_{(1,1)_{-1}}&=&{ M_{10}}+{ N_Y}
\end{array}
\right. \label{10dec}
 \ea
 Similarly, for the $\Sigma_{5}$ curves
 \ba {\bf 5}\in
SU(5)\Rightarrow\left\{\begin{array}{lll}
n_{(3,1)_{-\frac 13}}-n_{(\bar 3,1)_{\frac 13}}&=&{ M_5}\\
n_{(1,2)_{\frac 12}}-n_{(1, 2)_{-\frac 12}}&=&{ M_{5}}+{ N_Y}
\end{array}
\right. \label{5dec}
\ea

In the model under consideration $N_Y$ takes  values which are
combinations of $N_1, N_2$ as these are specified for the matter
curves in Table~\ref{SU5Z2Z2}, and similarly for $M_{10},M_5$
appearing in the next column. The resulting SM spectrum appears in
Table~\ref{SU5Z2Z2A}. Because of the  monodromies we have the
identifications $t_1=t_2$ and $t_3=t_4$. Therefore the $SU(5)$
relation $\sum_{i=1}^5t_i=0$ now becomes
\[ 2t_1+2 t_3+t_5=0\]
   \begin{table}[tbh] \centering%
\begin{tabular}{|l|c|c|l|}
\hline
 $SU(5)$& $U(1)_Y$&$U(1)_X$&SM spectrum\\\hline
$  10^{(1)}_{t_{1}}$&$ 0$ &$2$&$2\times (Q,u^c,e^c)$\\ $
10^{(2)}_{t_{3}}$&$ 1$ &$1$&$ (1\times Q,2\times e^c)$\\  $
10^{{(3)}}_{t_{5}}$&$-1$&$0$&$(1\times u^c,1\times \bar e^c)$\\
$ 5^{(0)}_{-2t_{1}}$&$ 0$ &$1$&$1\times (d, h_u)$\\
$\bar 5^{(1)}_{t_{1}+t_3}$&$ 0$ &$-1$&$1\times (d^c, \ell)$\\
 $ \bar 5^{(2)}_{t_{1}+t_5}$&$ 0$ &$-1$&$1\times (d^c, \ell)$\\
 $5^{(3)}_{t_{3}+t_5}$&$ -1$ &$0$&$1\times h_d$\\
 $ \bar 5^{(4)}_{2t_{3}}$&$1$&$-2$&$(2\times d^c, 1\times \ell)$\\
\hline
\end{tabular}%
\caption{Field representation content under $SU(5)$, their homology
class and flux restrictions under $U(1)_{Y}$ for the ${ Z}_2\times {
Z}_2$ case.} \label{SU5Z2Z2A}
\end{table}

Various low energy implications of the ${Z}_2\times { Z}_2$ models
have been discussed elsewhere~\cite{Dudas:2010zb,Antoniadis:2012yk}.
  Here we will focus only on the lepton sector and in particular on the neutrino mixing effects.

\subsubsection{Lepton Masses}

In this example we assume non-zero vevs to the singlet fields
$\theta_{13},\theta_{15},\theta_{35}$ and we define the ratios
\be\frac{\langle\theta_{13}\rangle}{M_X}=a,\; \frac{\langle\theta_{15}\rangle}{M_X}=b,\;
\frac{\langle\theta_{35}\rangle}{M_X}=c\label{abc}
\ee
where $M_X$ is the GUT scale.

 The $SU(5)$ Yukawa couplings of  the charged leptons are
 of the form
 \[ \lambda^{\ell}_{ij} \bar 5_i\,\bar 5_h\,10_j \]
 where  $\lambda^{\ell}_{ij}$ are coefficients calculated in terms of
 the integrals of the  wavefunctions corresponding to the states involved
 in the trilinear coupling.

  The three lepton doublets and down quark singlets $\bar f_i =(d^c,\ell)_i$ are accommodated
  according to
   \[ \bar f_1\in \bar 5_{t_3+t_4},\; \bar f_2\in \bar 5_{t_1+t_5},\;  \bar f_3\in \bar 5_{t_1+t_3}\]

Note that two RH-electrons $e^c$ reside on the same matter curve.
Further, there are two  more $e^c$'s and one $\bar e^c$. It is
therefore expected that one linear combination will form a massive
state  $\sim M_E\,\Sigma_i \alpha_i e_i\,\bar e^c$, while the
remaining three light degrees of freedom will correspond to those of
the three SM generations. Suppressing Yukawa coefficients and an
overall scale the charged lepton mass terms are
 \be
{\cal W}_{\ell}=\bar 5_{t_3+t_5} \left(\bar
5_{2t_3}(10_{t_3}\theta_{13}^2+10_{t_1}\theta_{13})+
\bar 5_{t_1+t_5}(10_{t_3}\theta_{15}+10_{t_1}\theta_{35}) +\bar 5_{t_1+t_3}(10_{t_3}\theta_{13}
+10_{t_1})\right)\nn
  \ee
  These terms will give rise to a charged lepton mass matrix of the form
  \[
M_{\ell}\approx \left(
\begin{array}
[c]{ccc}%
a^2 &\eps a  & a\\
b& \eps c&c\\
a&\eps &1
\end{array}
\right)
\]
The singlet vevs have been replaced here with the ratios $a,b,c$
defined in (\ref{abc}).
 The coefficient $\epsilon<1$ is introduced to account  for the flux effects
since two states are on the same matter curve. Families residing on
the identical matter curve may also be distinguished by phase
factors. We have also suppressed Yukawa coefficients $
\lambda^{\ell}_{ij}$. At tree-level these are calculated in terms of
integrals of overlapping wavefunctions of the three states
participating in the intersection. Higher order couplings are
mediated by appropriate KK-states, and therefore the Yukawa factors
are expected to be suppressed relative to the tree-level ones.
Therefore, we expect small mixing effects from the charged lepton
sector.

To construct the neutrino mass matrix, we note first that the three
left handed neutrinos are in the following fiveplets
\[ \nu_1\in \bar 5_{t_3+t_4},\; \nu_2\in \bar 5_{t_1+t_5},\;  \nu_3\in \bar 5_{t_1+t_3}\]
These should be coupled with RH-partners to generate Dirac type mass
terms
\[\lambda_{\nu}h_u\nu\nu^c\to m_D\nu\nu^c\]
RH-neutrinos couple among themselves with Majorana terms, while the
effective light Majorana neutrino mass matrix relevant to experiment
is obtained by the usual see-saw mechanism
\be
m_{\nu}^{eff}= -m_D M_{\nu^c}^{-1} m_D^T\label{seesaw}
\ee

For the RH neutrinos, we take the point of view of
reference~\cite{Antoniadis:2002qm} and identify them with the
Kaluza-Klein modes.  This mechanism has been implemented in F-theory
models~\cite{Bouchard:2009bu} and in our case operates as follows.
In the present model, the available singlet fields whose six
dimensional massive KK-modes would play the r\^ole of the
RH-neutrino, are $\theta_{ij}$, $i,j=1,\dots 5$.  For a particular
pair of indices $ij$ we then identify
\[\theta^{KK}_{ij}\to \nu^c , \theta^{KK}_{ji}\to \bar \nu^c\]
and the corresponding mass coupling is $M_{KK}   \nu^c\bar \nu^c$.
However, the above identifications imply that $ \nu^c$ and $\bar
\nu^c$ are complex representations transforming as  bifundamentals
in the intersections of
 the seven branes.  In general $ \nu^c$ and $\bar \nu^c$ cannot be identified.
For the particular cases of the  ${ Z}_2\times { Z}_2$
 monodromies, we have $t_1\leftrightarrow t_2 $
  and  $t_3\leftrightarrow t_4 $.  This implies $\theta_{12}=\theta_{21}$ and
$ \theta_{34}=\theta_{43}$. Thus, for these particular cases we
identify the matter curves and choose to interpret  the
corresponding KK-modes as the RH neutrino states
 \[\Sigma_a: \theta_{12}=\theta_{21}\to \nu^c_a ,\;\Sigma_b: \theta_{34}=\theta_{43}\to \nu^c_b\]
We accommodate the RH neutrino of the first generation on the matter
curve $\Sigma_b$ and the next two on $\Sigma_a$. We obtain the
following mass matrices (suppressing again Yukawa coefficients):

For the Dirac terms
\be
m_{D_{\nu}}=\left(
\begin{array}{ccc}
a^2&\epsilon a^2 &a^2\\
ac&\epsilon b&b\\
0&\pm\epsilon a&a
\end{array}
\right)m_{D_0}
\ee
and the heavy Majorana ones
\be
M_{{\nu}^c}=\left(
\begin{array}{ccc}
M&\epsilon a^2 M_X&a^2 M_X\\
\epsilon a^2 M_X&\epsilon^2 M&\epsilon M\\
a^2 M_X&\epsilon M&M'
\end{array}
\right) \ee In the Majorana case, we have introduced two different
$M_{KK} $ masses, of the same order $M\sim M'$ in correspondence
with the two neutrino species while $M_X$ stands for the GUT scale.

The see-saw matrix  is given by the known formula~(\ref{seesaw}).
Substitution of  the matrices $m_{D_{\nu}},M_{{\nu}^c}$ yields

\be m^{eff}_{\nu}= \left(
\begin{array}{ccc}
 2 a^4 & a^2 (b+a c) & a^3 \\
 a^2 (b+a c) & \frac{(c^2  a^2+b^2) M-2 b c a^3M_X }{M-a^2
  M_X} & \frac{a b M-a^4 c M_X}{M-a^2M_X} \\
 a^3 & \frac{a b M-a^4 c M_X}{M-a^2 M_X} & \frac{a^2 M}{M-a^2
   M_X}
\end{array}\right) m_{\nu_0}
\ee where $ m_{\nu_0}$ is an  effective light neutrino mass scale.

Since the above matrix is rather complicated, it is useful to
examine some limiting cases to check whether a large mixing
compatible with data is induced. Inasmuch the charged lepton mixing
is negligible, we  expect that the main source of the large neutrino
mixing comes from the neutrino mass matrix itself.

Let us take  the particular  limit $b\ll a,c$ and $M> a^2 M_X$. Then
we can approximate this matrix as follows
\be m^{eff}_{\nu} \approx
\left(
\begin{array}{ccc}
 2 a^2 & a (c+ r) & a \\
 a (c+ r) & c^2+r^2 & r \\
 a & r & 1
\end{array}
\right) \ee with $r=\frac{b}{a}$.

We have already assumed the case of small mixing in the charged
lepton sector, thus we would like to see whether large mixing
effects  eventually exist in the neutral part.  To get an idea, we
first set
 $r\to 0$ and assume $c\sim 1$. In this case the
diagonalizing matrix takes the form (setting $\tan\phi=\sqrt{2}\,a$)
\[ V_{\nu}=
\left(
\begin{array}{lll}
 \sin (\varphi ) & -\cos (\varphi ) & 0 \\
 \frac{\cos (\varphi )}{\sqrt{2}} & \frac{\sin (\varphi )}{\sqrt{2}} &
   -\frac{1}{\sqrt{2}} \\
 \frac{\cos (\varphi )}{\sqrt{2}} & \frac{\sin (\varphi )}{\sqrt{2}} &
   \frac{1}{\sqrt{2}}
\end{array}
\right)
\]
Observe now that for the value  $\tan\phi = \sqrt{2}$ this is just
the so called Tri-Bi maximal mixing
\[ \left(
\begin{array}{lll}
 -\sqrt{\frac{2}{3}} & -\frac{1}{\sqrt{3}} & 0 \\
 \frac{1}{\sqrt{6}} & -\frac{1}{\sqrt{3}} & -\frac{1}{\sqrt{2}} \\
 \frac{1}{\sqrt{6}} & -\frac{1}{\sqrt{3}} & \frac{1}{\sqrt{2}}
\end{array}
\right)\]

 Although this  very interesting result is achieved for extreme values of the parameters
 $a\sim c\sim 1$, it reveals that in the  ${ Z}_2\times { Z}_2$ model there is an intrinsic structure
 of the neutrino sector which incorporates in a natural way the
 large mixing effects as indicated by neutrino data. Therefore, below   we focus
 closer to  this promising case  and perform the analysis  restoring
 initially the non-zero value of the parameters $r=b/a$.  We define
 \[\xi =\frac{1}{1-\eta},\;  {\rm where}\; \eta
 =\frac{M_X}{M}a^2=\frac{\langle\theta_{13}\rangle}{M}\frac{\langle\theta_{13}\rangle}{M_X}\]
Then the effective neutrino mass matrix can be cast to the
convenient form

\be
m^{eff}_{\nu} \approx \left(
\begin{array}{ccc}
 2 a^2 & a (c+ r) & a \\
 a (c+ r) & (c-r)^2 \xi+2cr & c-(c-r)\xi \\
 a &  c-(c-r) \xi& \xi
\end{array}
\right)
\ee
In this limit, the light Majorana mass matrix has one zero
eigenvalue. To check whether we obtain a reasonable parameter space,
we write down the ratio of the mass square differences
\[ \Delta m^2_{sun}=|m_2^2-m_1^2|,\; \Delta
m^2_{atm}=|m_3^2-m_2^2|\] which is experimentally measured to be
 \be
 \frac{\Delta m^2_{atm}}{\Delta m_{sun}^2}=30\label{M2ratio}
 \ee

\begin{figure}[!t]
\centering
\includegraphics[scale=0.7]{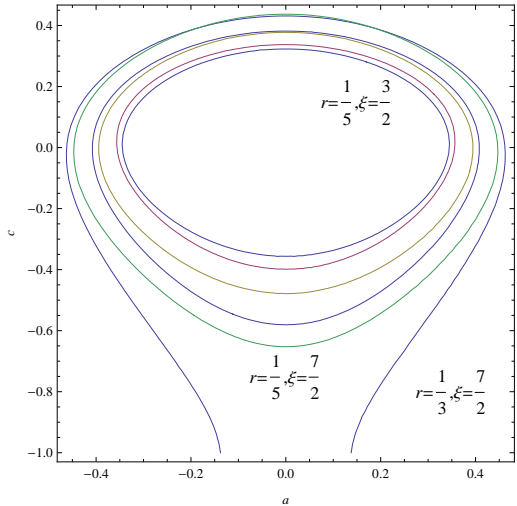}
\caption{Curves  for the ratio $\Delta m_{32}^2/\Delta m_{31}^2=30$
in the parameter space $(a,c)$.} \label{Dm2_ac}
\end{figure}

In figure~\ref{Dm2_ac} we plot contours of the above ratio in the
plane $(a,c)$ for  various values of  the pairs and in particular
\[ (r,\xi)=(\frac 15,\frac 32),\;(\frac 13,\frac 52),\;
(\frac 13,\frac 32),\;(\frac 15,\frac 52),\;(\frac 15,\frac
72),\;(\frac 13,\frac 72)\] We observe that the experimentally
measured value of this ratio is obtained for reasonable range of the
parameters $a,b,r,\xi$.

Having checked that the parameters $a,c$ are in the perturbative
range, while consistent with the mass data, we proceed to the mixing
matrix. We can find easily numerical solutions for a wide range of
the parameters consistent with the  experimental data.  In our
approach, we have assumed a negligible mixing in the charged lepton
sector therefore, the large mixing effects of the lepton mixing
matrix $V=V_{\ell}^{\dagger}V_{\nu}$ relevant to experiment, are
expected from the neutral sector.

Therefore, for our present purposes we assume almost diagonal
$V_{\ell}$ and  only consider the neutrino mixing matrix. Moreover,
to simplify further the analysis in the present application, we will
reduce the parameter space by imposing the condition $c+r=1$. This
simplification is motivated by the most general structure of the the
TB-mixing matrix in~(\ref{mnTB}) where the relation $m_{12}=m_{13}$
holds. Of course in this way we restrict our investigations to a
smaller portion of the full parameter space, but our purposes here
is to only show that large mixing effects consistent with
experiments are naturally accommodated in these models.

\begin{figure}[!t`]
\centering
\includegraphics[scale=0.9]{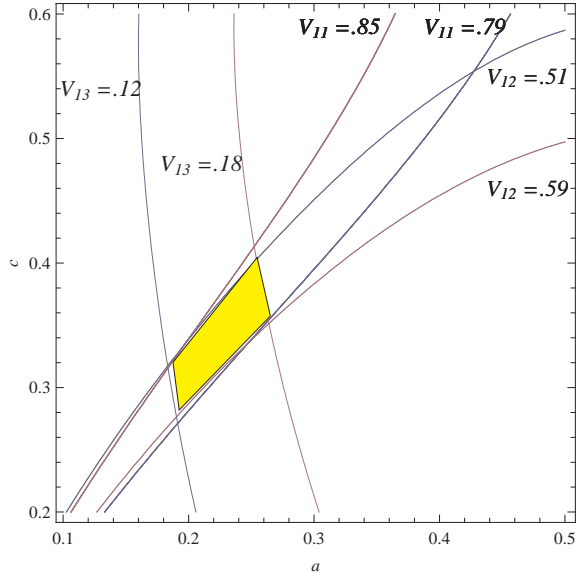}
\caption{ Bounds  in the parameter space $(a,c)$ from the
experimental range of neutrino mixing $V_{\nu}$ elements.}
\label{Vranges}
\end{figure}

 We use the ranges of the elements $V_{ij}$ as they are determined by
 the present-day experiments and vary  $a,c$ to see if there are consistent values of the latter.
 We first observe that the elements $V_{23},V_{33}$ allow a wide range of values.
 Significant restrictions arise only from the elements $V_{11},V_{12},V_{13}$.
 In figure \ref{Vranges} we plot the bounds put by the experimental
 ranges~\cite{GonzalezGarcia:2012sz}.  In particular we plot the neutrino mixing entries for the following ranges
\[ V_{11}=[0.795-0.846], V_{12}=[0.513-0.585], V_{13}=[]0.126-0.178], V_{22}=[0.416-0.730]\]

We notice that indeed, there exists an overlapping region where all
the constraints imposed by the $V_{ij}$ ranges dictated by the
neutrino experiments are satisfied. Furthermore,  the $a,c$
parameters take values $<1$  in consistency with the requirements
that the singlet vevs should be in the perturbative regime
$\langle\theta_{ij}\rangle < M_X$. Note however, that in this
restricted region  the mass-square differences ratio comes out to be
larger than the experimental value (\ref{M2ratio}), but
renormalization group effects are expected to change this in any
case.

\section{Conclusions}

In this work we have investigated  the neutrino properties in local
F-theory GUT models which are realised on singular  CY four-folds.
We focused on the minimal unified symmetry $SU(5)_{GUT}$ embedded in
the exceptional group $E_8$ in the elliptic fibration, and   the
properties emerging from its associated spectral cover ${\cal C}_5$
possessing an $SU(5)_{\perp}$ symmetry. To reproduce known neutrino
properties such as large mixing which is nearly maximal among two
generations, we consider the splitting
 ${\cal C}_5\to {\cal C}_4\times {\cal C}_1$ of  spectral cover and study
all possible monodromies `disguised'  as surrogate discrete
symmetries in the low energy effective theory limit. We worked out
constraints on the fermion mass structure with respect to the
monodromies of the ${\cal C}_4$ part, and in particular to those
associated to the permutation symmetries $S_4, A_4$ as well as
several of its subgroups. We paid particular attention to the
leptonic sector and derived the charged lepton and neutrino mass
textures for two models based on $A_4$ and ${Z}_2\times { Z}_2$
family symmetries. In the $A_4$ model right-handed neutrinos are not
introduced and consequently the effective neutrino mass matrix is
assumed to arise from a fifth order non-renormalisable operator. It
is found that such a symmetry imposes significant restrictions on
the F-theory models since  there is only a limited number of matter
curves to accommodate fermion families. Interestingly, however, the
study of the leptonic sector shows that large mixing effects
compatible with the recent neutrino data can be accommodated
naturally. The second model we have examined is based on a ${
Z}_2\times  {Z}_2$ family symmetry. In this model the lepton sector
includes also right handed neutrinos represented by appropriate
massive KK states. This model also induces a large lepton mixing
matrix in a natural way. Thus, we found that large lepton mixing
effects  are a generic property of these F-theory constructions. In
particular, this property is linked directly to the neutral part of
the lepton sector, while the charged lepton mixing matrix in both
cases is found to be  small.

\vfill {\bf Acknowledgements:} {\it This research has been
co-financed by the European Union (European Social Fund - ESF) and
Greek national funds through the Operational Program "Education and
Lifelong Learning" of the National Strategic Reference Framework
(NSRF) - Research Funding Program: "ARISTEIA". Investing in the
society of knowledge through the European Social Fund. This work was
supported in part by the European Commission under the ERC Advanced
Grant 226371 and the contract PITN-GA-2009- 237920. GKL would like
to thank Theory Division, CERN for kind hospitality.}

\newpage

\appendix{\bf Appendix}

We present here several computations and details with regard to the discriminant
$\Delta(a_i)$ of the fourth degree polynomial associated to ${\cal
C}_4$-spectral cover. Since we express $\Delta(a_i)$  as a
polynomial with respect to one of its parameters $a_i$, we first
start with some properties of the factorisation.

\section{The factorisation of the polynomial}

We are interested in factorisations of the polynomial whose
coefficients $c_k$ belong to a definite field ${\cal F}$.  Consider the general fourth-degree one
\[P_4=c_4 x^4+c_3 x^3+c_2x^2+c_1x+c_0\]
with $c_4>0$.  We wish to write the polynomial as a square
\[ P_4= ( b_2 x^2+b_1 x+b_0)^2\]
This implies
\[ b_2=\sqrt{c_4},\;b_0=\sqrt{ c_0}\]
and various  relations among the coefficients.

Since we assume  $c_k \in {\cal F}$  and we also  want $b_k\in {\cal
F}$ we readily conclude that $\sqrt{c_4}$ and $\sqrt{c_0}$ must be
in ${\cal F}$ too:
\[ \sqrt{c_4},\sqrt{c_0}\in {\cal F}\]

Next, assume a fourth-degree polynomial of the particular form
\be
P_4(x)=
(x^2+2\kappa x+\lambda^2)^2-\mu^4 \label{P4sep}
\ee
This can split to two factors
\[P_4(x)= (x^2+2\kappa x+\lambda^2+\mu^2)(x^2+2\kappa
x+\lambda^2-\mu^2)\]
while it is automatically a square when $\mu=0$.

Computing the discriminant we find
\be
\Delta_{P_4}=256 \mu ^8 \left(\left(\kappa ^2-\lambda
^2\right)^2-\mu ^4\right)
\ee
Therefore, when the discriminant is zero the following possibilities
emerge
\ba
\Delta_{P_4}=0\lra \left\{\begin{array}{lcl}
                    \mu=0& \Rightarrow&P_4 =\delta^2\\
                    \mu^2=|\kappa^2-\lambda^2|&\Rightarrow&P_4=
                    \delta_1\,\delta_2
                    \end{array}
                    \right.
                    \label{Dcases}
                    \ea
with
\ba
\delta&=&(x^2+2\kappa x+\lambda^2)\\
\delta_1&=&(x+\kappa)^2\\
\delta_2&=&x^2+2\kappa x+2\lambda^2-\kappa^2
\ea
Unless further relations are imposed among $\kappa, \lambda, \mu$, we observe that only one of the two zeros of the discriminant
implies $P_4= \delta^2$, the other just giving a simple factorisation $P_4=\delta_1\delta_2$.

\section{The Discriminant}

 The starting point to achieve an  $S_4$ discrete subgroup such as $A_4$
  is to study the conditions on the  factorisation of the discriminant and for $A_4$
  is particular to write it as a square
 $\Delta =\delta^2$. In this appendix we will use  the  following
 $b_1=0$ condition
\[\left\{a_4\to -a_0 a_6,a_5\to a_0 a_7\right\}\]
and substitute to the determinant  $S_d$ in (\ref{Disc0}). Then
$S_d$ takes the form
   \be
   \label{Discmin}
\begin{split}
S_d=& a_0^2 a_7 \left(-27 a_0^3 a_1^2 a_6^4+4 a_3^3 \left(4 a_1
a_3-a_2^2\right) a_7+a_0 \left(a_3^2
   \left(a_2^2-4 a_1 a_3\right) a_6^2\right.\right.\\
   &\left.\left.+2 a_2 a_3 \left(40 a_1 a_3-9 a_2^2\right) a_7 a_6+\left(-27
   a_2^4+144 a_1 a_3 a_2^2-128 a_1^2 a_3^2\right) a_7^2\right)\right.\\
   &\left.+2 a_0^2 \left(a_2 \left(2 a_2^2-9
   a_1 a_3\right) a_6^3-3 a_1 \left(a_2^2-24 a_1 a_3\right) a_7 a_6^2+96 a_1^2 a_2 a_7^2 a_6+128
   a_1^3 a_7^3\right)\right)
    \end{split}
   \ee
Then, the discriminant is given by
\[\Delta=\frac{1}{a_5}S_d=\frac{1}{a_0a_7}S_d\]

 We wish to investigate under what conditions this is separable with
factors expressed in terms of coefficients $\in {\cal F}$. Noticing
that $\Delta$ can be expressed as a polynomial of a given
coefficient $P(a_k)$, according to the discussion above we can
factorise $P(a_k)$ by imposing the condition that its own
discriminant is zero.  Expanding  the discriminant as a polynomial
of the various coefficients $a_k$ we obtain
\be
\begin{split}
f_5&=0 \,{\bf a_5^4}+256 a_1^3 {\bf a_5^3}+\cdots
\\
f_4&=-27 a_1^2 {\bf a_4^4}+\cdots
\\
f_3&=16 a_1 a_5{\bf a_3^4 }+\cdots
\\
f_2&= -27  a_5^2{\bf a_2^4}+\cdots
\\
f_1&=0 \,{\bf a_1^4}+256 a_5^3{\bf  a_1^3}+\cdots
\end{split}
\ee

A few remarks are in order:

\begin{enumerate}

\item

 For real $a_1,a_5$,  the  polynomials  $f_2,f_4$ have a negative coefficient, therefore cannot be positive definite.

 \item

$f_5, f_1$ are cubic polynomials and  and a necessary  condition to be positive definite is
  sign$({a_1a_5})=(+)$.

 \item The  remaining possibility is $f_3$ is also positive definite
 for sign$({a_1a_5})=(+)$.  We will choose to work out this case for
 the reasons discussed in the text.

 \item

This polynomial has the general form
\[f_3=c_4 a_3^4+c_3 a_3^3+c_2a_3^2+c_1a_3+c_0\]
($c_4>0$) or written as a square:
\[ f_3= ( b_2 a_2^2+b_1 a_2+b_0)^2\]
We have just seen that   $ b_2=\sqrt{c_4},\;b_0=\sqrt{ c_0}$ and
 since we assume  $c_k \in {\cal F}$  and we also  want $b_k\in
{\cal F}$ we readily conclude that $\sqrt{c_4}$ and $\sqrt{c_0}$
must be in ${\cal F}$ too:
\[ \sqrt{c_4},\sqrt{c_0}\in {\cal F}\]
In our case, for $c_4$ we must have
\[ c_4\sim a_1 a_5= a_0a_1a_7 = e_3^2, \;\; e_3\in  {\cal F}\]
while  $c_0$ has a complicated form. Instead, we can extract some
useful relations from the discriminant.

 \end{enumerate}

\subsection{{Analysis of $f_3$ case}}

Computing the discriminant of $f_3$ , we find that it is
proportional to  $\Delta_{f_3}\propto D_a^2D_b^3$ with
\be
\label{DaDb}
\begin{split}
D_a&=a_1 a_4^2-a_2^2 a_5\\
D_b&=\left(27 a_1^2 a_4-a_2^3\right) a_4^3-6 a_1 a_2^2 a_5
a_4^2+4096 a_1^3 a_5^3+3
   a_2 \left(9 a_2^3-256 a_1^2 a_4\right) a_5^2
   \end{split}
   \ee
Therefore, the discriminant is zero either when $D_a=0$ or if
$D_b=0$.

We will see that, in accordance  with what we have seen above, one
condition, namely $D_a=0$, leads to a factorisation of the
polynomial $f_3(a_3)\to \Delta\,  =\, \delta_1\delta_2$ while to
succeed writing  $f_3(a_3)\to \Delta\,  =\, \delta^2$  further
conditions should be met. Next, we will see what conditions emerge
from the second case.

\subsubsection{ $D_a=0$ case}

We combine now the $a_1a_5=e_3^2$ condition with $D_a=0$:
\be
a_1 a_4^2=a_2^2 a_5\label{Dazero}
\ee
 From $a_1 D_a=0$
\[ (a_1a_4)^2 =a_2^2 a_1a_5 =  (a_2e_3)^2\]
thus   $a_1a_4=\pm e_3a_2$. From $a_5 D_a=0$   we also get
$a_2a_5=\pm e_3a_4$. Therefore, summarising, we end up with the
following set of relations
\be
\label{Dac4}
\begin{split}
a_1a_4&=\pm e_3a_2\\
a_2a_5&=\pm e_3a_4\\
a_1a_5&=e_3^2
\end{split}
\ee
To correlate the signs we multiply the second by $a_1$ and get
$a_2a_1a_5=\xi\, e_3a_1a_4$ with $\xi=\pm 1$. We substitute $a_1a_5
=e_3^2$ in the first term to obtain $e_3^2a_5=\xi\, e_3a_1a_4$ or
$a_2e_3=\xi\, a_1a_4$.

The polynomial with respect to $a_3$  is
\[f_3= c_4a_3^4+c_3a_3^3+c_2a_3^2+c_1a_3^1+c_0\]
with coefficients
\be
\label{P4a3}
\begin{split}
c_4&=16 a_1 a_5\\
c_3&=-4(a_5 a_2^2+a_1 a_4^2)\\
c_2&=a_2^2 a_4^2-80a_2 a_4 a_1 a_5 -128 a_1^2 a_5^2\\
c_1&=18 a_4 a_5 a_2^3+144 a_1 a_5^2 a_2^2+18 a_1 a_4^3 a_2+144 a_1^2 a_4^2 a_5\\
c_0&=-\left(4 a_2^3+27 a_1^2 a_4\right) a_4^3-6 a_1 a_2^2 a_5
a_4^2+256 a_1^3
   a_5^3-3 a_2 \left(9 a_2^3+64 a_1^2 a_4\right) a_5^2
   \end{split}
   \ee
Now we use the relations
\be
 a_4=\xi e_3 a_2,\; a_1a_5=e_3^2\label{a1245}
 \ee
whilst we introduce the redefinitions,
\be
e_3a_1\to  \beta_2^2,\;  a_1a_3 \to  x\label{a1a3x}
\ee
so that the polynomial $f_3$ takes the form
\[
16 x^4-8a_2^2x^3+(a_2^4-80 a_2^2 \beta _2^2-128 \beta _2^4)x^2+36
a_2^2 \beta _2^2 \left(a_2^2+8 \beta _2^2\right)x+ 4 \beta _2^2
\left(\beta _2^2-a_2^2\right) \left(a_2^2+8 \beta
   _2^2\right)^2
\]
Clearly this cannot be automatically a square, however, it is easy
to notice that this case corresponds to the second case of
(\ref{Dcases}), i.e., when $\Delta_{f_3}=\delta_1\delta_2$. Indeed,
after some algebra we find
\be
   \label{factorf3}
\begin{split}
f_3&=\left(a_2^2+8  \beta_2^2-4 x\right){}^2 \left(\left(2  \beta_2^2+x\right)^2-4
   a_2^2  \beta_2^2\right)
\end{split}
\ee
Now, we impose the relation
\be
2x\to 2a_1a_3 =a_2^2\label{xa3}
\ee
Substitution of the above  in (\ref{factorf3}) yields

\be
   \label{factorff3}
\begin{split}
f_3&=\left(a_2^2+8  \beta_2^2-2a_2^2\right){}^2 \left(\left(2  \beta_2^2+a_2^2/2\right)^2-4
   a_2^2  \beta_2^2\right)\\
   &=\frac{1}{4}\left( \left(8  \beta_2^2-a_2^2\right) \left(4 \beta_2^2-a_2^2\right)\right)^2\\
   &= \frac{1}{4}\left[ a_1^2\left(8  \,e_3-a_3\right) \left(4 \,e_3-a_3\right)\right]^2
\end{split}
\ee
where we have used  consecutively the  relations in
(\ref{a1245}) and (\ref{a1a3x}).  Hence,
we have succeeded to write   $f_3\to \Delta_{f_3}=\delta^2$ with
\be
\delta= \frac{1}{2}\, a_1^2\left(8  \,e_3-a_3\right) \left(4 \,e_3-a_3\right)
\ee

If we combine the two relations (\ref{Dazero}) and (\ref{a1a3x}) we find
\be
\label{Dafinal}
\begin{split}
a_2^2&=2\,a_1\,a_3\\
a_4^2&=2\,a_3\,a_5
\end{split}
\ee
Using (\ref{Dafinal}) we rewrite $\Delta=\delta^2$ only in terms of the initial coefficients $a_k$:
\be
a_3^6 \Delta \;= \;\left(a_3^3\delta\right)^2\;=\;\left[a_2 a_4 \left(a_3^2-2\, a_2 a_4\right) \left(a_3^2-a_2 a_4\right)\right]^2
\ee

In the subsequent analysis  we
examine the case  $D_b=0$ which also leads to a simple  condition
among the coefficients $a_k$.

\subsubsection{Factorisation of matter curves}

One of the main goals is to use these relations to factorise
appropriately the matter curves. For the  case of the $\Sigma_5$
curve, we have seen that when the symmetry is $S_4$ it splits to two
pieces $P_5=P_5^aP_5^b$. Further splitting can occur  using the
$D_a=0$ relation.  For example,  multiplying $a_0 P_5^a$ we have
\be
\begin{split}
a_0 P_5^a& =a_0 (a_2^2a_7+ a_2a_3 a_6+ a_0 a_1 a_6^2)\\
                &=a_2^2(a_0a_7)+a_2a_3(a_0a_6)+a_1(a_0a_6)^2\\
                &=a_2^2 a_5-a_2a_3a_4+a_1a_4^2\\
                &=a_1a_4^2-a_2a_3a_4+a_1a_4^2\\
                &=a_4(2a_1a_4-a_2a_3)\\
                &=a_0a_6(2a_0a_1a_6-a_2a_3)
\end{split}
\ee
where we have used successively the $b_1=0$ and $D_a=0$ conditions.
Eliminating $a_0$ from the first and last lines of the above, we get
\[P_5^a=a_6(2a_0a_1a_6-a_2a_3)\]
Therefore the fiveplets' equation  separates now to three distinct
curves
\[ P_5=a_6(2a_0a_1a_6-a_2a_3)  \left(a_3 a_6^2+
\left(a_2 a_6+a_1  a_7\right)a_7\right)\]

\subsubsection{The $D_b=0$ case}

The vanishing of the discriminant $\Delta_{f_3}$ can happen also
when  the second factor in (\ref{DaDb}) is set to zero, i.e.  $D_b=0$,  which leads to
analogous relations among $a_k$. Indeed, it can be readily checked
that this factor can be brought to the form
\be
\label{Db1}
\begin{split}
D_b&= (16)^3 (a_1a_5)^3\\
&-3\cdot (16)^2 (a_1a_5)^2 (a_2a_4)\\
       &+27 (a_2^4a_5^2+a_4^4a_1^2)\\
       &-6  (a_1a_5)(a_2a_4)^2\\
       &-(a_2a_4)^3
\end{split}
\ee
Observing the third term, we can easily recognise that conditions
(\ref{Dac4})  transform this term to
\[ (a_2^4a_5^2+a_4^4a_1^2) = 2 (e_3a_2a_4)^2 \equiv 2 (a_1a_5) (a_2a_4)^2\]
Substituting to  (\ref{Db1}) we get
\be
D_b = (16 a_1a_5-a_2a_4)^3\label{Dbcube}
\ee
This could be solved for a particular relation among the
coefficients, namely $a_2a_4 =16 a_1a_5\sim 16\,e_3^2$ however,  we
implicitly used the condition $D_a=0$ which overconstrains the
system of the coefficients.

In fact, the requirement to have the condition $\Delta = \delta^2$
is more involved and the general solution should be found without
the implicit use of the result of the first case.

Indeed, we can solve the $D_b=0$ directly. Introducing $x=a_1a_5,
y=a_2a_4$, we get
\be
\begin{split}
27 (a_2^4a_5^2+a_4^4a_1^2)&=y^3+6 x y^2+3 (16x)^2y-(16x)^3
\\
&=(y-16x)^3+54 xy^2
\end{split}
\ee
Therefore the relation among the coefficients implying $\Delta
=\delta^2$ is
\ba
 a_2^4a_5^2+a_4^4a_1^2&=&\left(\frac{16a_1a_5-a_2a_4}{3}\right)^3+2 (a_1a_5)(a_2a_4)^2
\ea
Rearranging the term  $2 (a_1a_5)(a_2a_4)^2= 2 (a_1a_4^2)(a_5^2a_2)$
and moving it to the left-hand side of
 the latter equation we notice  that it can be further simplified
and written as an ``elliptic curve''  formula  $Y^2=X^3$:
\ba
( a_2^2a_5-a_4^2a_1)^2&=&\left(\frac{16a_1a_5-a_2a_4}{3}\right)^3
\ea
Plugging in the condition $b_1=0$ yields
\ba
(
a_2^2a_7-a_0a_1a_6^2)^2&=&a_0\left(\frac{16a_1a_7+a_2a_6}{3}\right)^3
\ea


\newpage
\begin{thebibliography}{99}

\bibitem{Beasley:2008dc}
  C.~Beasley, J.~J.~Heckman and C.~Vafa,
  ``GUTs and Exceptional Branes in F-theory - I,''
  JHEP {\bf 0901} (2009) 058
  [arXiv:0802.3391 [hep-th]].

\bibitem{Donagi:2008kj}
  R.~Donagi and M.~Wijnholt,
  ``Breaking GUT Groups in F-Theory,''
  Adv.\ Theor.\ Math.\ Phys.\  {\bf 15} (2011) 1523
  [arXiv:0808.2223 [hep-th]].

\bibitem{Beasley:2008kw}
  C.~Beasley, J.~J.~Heckman and C.~Vafa,
  ``GUTs and Exceptional Branes in F-theory - II: Experimental Predictions,''
  JHEP {\bf 0901} (2009) 059
  [arXiv:0806.0102 [hep-th]].

        \bibitem{Donagi:2008ca}
          R.~Donagi and M.~Wijnholt,
          ``Model Building with F-Theory,''
          Adv.\ Theor.\ Math.\ Phys.\  {\bf 15} (2011) 1237
          [arXiv:0802.2969 [hep-th]].

          \bibitem{Donagi:2009ra}
            R.~Donagi and M.~Wijnholt,
            ``Higgs Bundles and UV Completion in F-Theory,''
            arXiv:0904.1218 [hep-th].


\bibitem{Heckman:2010bq}
  J.~J.~Heckman,
  ``Particle Physics Implications of F-theory,''
  Ann.\ Rev.\ Nucl.\ Part.\ Sci.\  {\bf 60} (2010) 237
  [arXiv:1001.0577 [hep-th]].

\bibitem{Weigand:2010wm}
  T.~Weigand,
  ``Lectures on F-theory compactifications and model building,''
  Class.\ Quant.\ Grav.\  {\bf 27} (2010) 214004
  [arXiv:1009.3497 [hep-th]].

\bibitem{Leontaris:2012mh}
  G.~K.~Leontaris,
  ``Aspects of F-Theory GUTs,''
  PoS CORFU {\bf 2011} (2011) 095
  [arXiv:1203.6277 [hep-th]].


\bibitem{Maharana:2012tu}
  A.~Maharana and E.~Palti,
  ``Models of Particle Physics from Type IIB String Theory and F-theory: A Review,''
  Int.\ J.\ Mod.\ Phys.\ A {\bf 28} (2013) 1330005
  [arXiv:1212.0555 [hep-th]].


\bibitem{Cecotti:2009zf}
  S.~Cecotti, M.~C.~N.~Cheng, J.~J.~Heckman and C.~Vafa,
  ``Yukawa Couplings in F-theory and Non-Commutative Geometry,''
  arXiv:0910.0477 [hep-th].

\bibitem{Font:2009gq}
  A.~Font and L.~E.~Ibanez,
  ``Matter wave functions and Yukawa couplings in F-theory Grand Unification,''
  JHEP {\bf 0909} (2009) 036
  [arXiv:0907.4895 [hep-th]].

\bibitem{Font:2012wq}
  A.~Font, L.~E.~Ibanez, F.~Marchesano and D.~Regalado,
  ``Non-perturbative effects and Yukawa hierarchies in F-theory SU(5) Unification,''
  JHEP {\bf 1303} (2013) 140
  [arXiv:1211.6529 [hep-th]].


\bibitem{Dudas:2009hu}
  E.~Dudas and E.~Palti,
  ``Froggatt-Nielsen models from E(8) in F-theory GUTs,''
  JHEP {\bf 1001} (2010) 127
  [arXiv:0912.0853 [hep-th]].

\bibitem{King:2010mq}
  S.~F.~King, G.~K.~Leontaris and G.~G.~Ross,
  ``Family symmetries in F-theory GUTs,''
  Nucl.\ Phys.\ B {\bf 838} (2010) 119
  [arXiv:1005.1025 [hep-ph]].



\bibitem{Dudas:2010zb}
  E.~Dudas and E.~Palti,
  ``On hypercharge flux and exotics in F-theory GUTs,''
  JHEP {\bf 1009} (2010) 013
  [arXiv:1007.1297 [hep-ph]].

\bibitem{Leontaris:2010zd}
  G.~K.~Leontaris and G.~G.~Ross,
  ``Yukawa couplings and fermion mass structure in F-theory GUTs,''
  JHEP {\bf 1102} (2011) 108
  [arXiv:1009.6000 [hep-th]].

\bibitem{Camara:2011nj}
  P.~G.~Camara, E.~Dudas and E.~Palti,
  ``Massive wavefunctions, proton decay and FCNCs in local F-theory GUTs,''
  JHEP {\bf 1112} (2011) 112
  [arXiv:1110.2206 [hep-th]].

\bibitem{Callaghan:2011jj}
  J.~C.~Callaghan, S.~F.~King, G.~K.~Leontaris and G.~G.~Ross,
  ``Towards a Realistic F-theory GUT,''
  JHEP {\bf 1204} (2012) 094
  [arXiv:1109.1399 [hep-ph]].
  \\
  J.~C.~Callaghan and S.~F.~King,
  ``E6 Models from F-theory,''
  JHEP {\bf 1304}, 034 (2013)
  [arXiv:1210.6913 [hep-ph]].



  \bibitem{Antoniadis:2012yk}
    I.~Antoniadis and G.~K.~Leontaris,
    ``Building SO(10) models from F-theory,''
    JHEP {\bf 1208} (2012) 001
    [arXiv:1205.6930 [hep-th]].

\bibitem{Pawelczyk:2013tza}
  J.~Pawelczyk,
  ``A F-GUT inspired model of Yukawa couplings with matter-messenger unification,''
  arXiv:1305.5162 [hep-ph].

\bibitem{deMedeirosVarzielas:2006fc}
  I.~de Medeiros Varzielas, S.~F.~King and G.~G.~Ross,
  ``Neutrino tri-bi-maximal mixing from a non-Abelian discrete family symmetry,''
  Phys.\ Lett.\ B {\bf 648} (2007) 201
  [hep-ph/0607045].
  \\
  G.~Altarelli and F.~Feruglio,
  ``Discrete Flavor Symmetries and Models of Neutrino Mixing,''
  Rev.\ Mod.\ Phys.\  {\bf 82} (2010) 2701
  [arXiv:1002.0211 [hep-ph]].
  \\
  H.~Ishimori, T.~Kobayashi, H.~Ohki, Y.~Shimizu, H.~Okada and M.~Tanimoto,
  ``Non-Abelian Discrete Symmetries in Particle Physics,''
  Prog.\ Theor.\ Phys.\ Suppl.\  {\bf 183} (2010) 1
  [arXiv:1003.3552 [hep-th]].
  \\
  S.~F.~King and C.~Luhn,
  ``Neutrino Mass and Mixing with Discrete Symmetry,''
  Rept.\ Prog.\ Phys.\  {\bf 76} (2013) 056201
  [arXiv:1301.1340 [hep-ph]].
  \\
    R.~N.~Mohapatra, S.~Antusch, K.~S.~Babu, G.~Barenboim, M.~-C.~Chen, A.~de Gouvea, P.~de Holanda and B.~Dutta {\it et al.},
    ``Theory of neutrinos: A White paper,''
    Rept.\ Prog.\ Phys.\  {\bf 70} (2007) 1757
    [hep-ph/0510213].


\bibitem{Ibanez:1991hv}
  L.~E.~Ibanez and G.~G.~Ross,
  ``Discrete gauge symmetry anomalies,''
  Phys.\ Lett.\ B {\bf 260} (1991) 291.
\\
  P.~Anastasopoulos, M.~Cvetic, R.~Richter and P.~K.~S.~Vaudrevange,
  ``String Constraints on Discrete Symmetries in MSSM Type II Quivers,''
  JHEP {\bf 1303} (2013) 011
  [arXiv:1211.1017 [hep-th]].
  \\
  H.~M.~Lee, S.~Raby, M.~Ratz, G.~G.~Ross, R.~Schieren, K.~Schmidt-Hoberg and P.~K.~S.~Vaudrevange,
  ``A unique $Z_4^R$ symmetry for the MSSM,''
  Phys.\ Lett.\ B {\bf 694} (2011) 491
  [arXiv:1009.0905 [hep-ph]].
  \\
  L.~E.~Ibanez, A.~N.~Schellekens and A.~M.~Uranga,
  ``Discrete Gauge Symmetries in Discrete MSSM-like Orientifolds,''
  Nucl.\ Phys.\ B {\bf 865} (2012) 509
  [arXiv:1205.5364 [hep-th]].

\bibitem{BerasaluceGonzalez:2012vb}
  M.~Berasaluce-Gonzalez, P.~G.~Camara, F.~Marchesano, D.~Regalado and A.~M.~Uranga,
  ``Non-Abelian discrete gauge symmetries in 4d string models,''
  JHEP {\bf 1209} (2012) 059
  [arXiv:1206.2383 [hep-th]].


\bibitem{Marsano:2009gv}
  J.~Marsano, N.~Saulina and S.~Schafer-Nameki,
  ``Monodromies, Fluxes, and Compact Three-Generation F-theory GUTs,''
  JHEP {\bf 0908} (2009) 046
  [arXiv:0906.4672 [hep-th]].

\bibitem{Tate75}
J. Tate, ``Algorithm for Determining the Type of a Singular Fiber in
an Elliptic Pencil,''  in Modular Functions of One Variable IV,
Lecture Notes in Math. vol. 476, Springer-Verlag, Berlin (1975).



\bibitem{Choi:2010nf}
  K.~-S.~Choi,
  ``SU(3) x SU(2) x U(1) Vacua in F-Theory,''
  Nucl.\ Phys.\ B {\bf 842}, 1 (2011)
  [arXiv:1007.3843 [hep-th]].


\bibitem{Hayashi:2009bt}
  H.~Hayashi, T.~Kawano, Y.~Tsuchiya and T.~Watari,
  ``Flavor Structure in F-theory Compactifications,''
  JHEP {\bf 1008} (2010) 036
  [arXiv:0910.2762 [hep-th]].



\bibitem{Marsano:2012yc}
  J.~Marsano, H.~Clemens, T.~Pantev, S.~Raby and H.~-H.~Tseng,
  ``A Global SU(5) F-theory model with Wilson line breaking,''
  JHEP {\bf 1301} (2013) 150
  [arXiv:1206.6132 [hep-th]].




        \bibitem{Artin}
        Michael Artin,
        "Algebra",   Prentice-Hall Inc.  1991.
        \\
        Patrick Morandi, "Field and Galois Theory",  Springer, 1996.






      \bibitem{Luhn:2007sy}
        C.~Luhn, S.~Nasri and P.~Ramond,
        ``Tri-bimaximal neutrino mixing and the family symmetry semidirect product of Z(7) and Z(3),''
        Phys.\ Lett.\ B {\bf 652} (2007) 27
        [arXiv:0706.2341 [hep-ph]].





\bibitem{Antoniadis:2002qm}
  I.~Antoniadis, E.~Kiritsis, J.~Rizos and T.~N.~Tomaras,
  ``D-branes and the standard model,''
  Nucl.\ Phys.\ B {\bf 660} (2003) 81
  [hep-th/0210263].





     \bibitem{Bouchard:2009bu}
        V.~Bouchard, J.~J.~Heckman, J.~Seo and C.~Vafa,
        ``F-theory and Neutrinos: Kaluza-Klein Dilution of Flavor Hierarchy,''
        JHEP {\bf 1001} (2010) 061
        [arXiv:0904.1419 [hep-ph]].



\bibitem{GonzalezGarcia:2012sz}
  M.~C.~Gonzalez-Garcia, M.~Maltoni, J.~Salvado and T.~Schwetz,
  ``Global fit to three neutrino mixing: critical look at present precision,''
  JHEP {\bf 1212} (2012) 123
  [arXiv:1209.3023 [hep-ph]].




  \end{thebibliography}
\end{document}